\documentclass[twocolumn,aps,pra,superscriptaddress,nobibnotes,amsmath,amssymb,floatfix,footinbib,longbibliography]{revtex4-2}
\usepackage[markup=nocolor, authormarkupposition=left]{changes}
\usepackage{soul}
\usepackage[utf8]{inputenc}
\usepackage{amsmath,amsfonts,amssymb}
\usepackage[T1]{fontenc}
\usepackage{url}
\usepackage{xcolor}
\usepackage{siunitx}
\usepackage{epstopdf}
\usepackage{graphicx}
\usepackage{changes}
\usepackage{float}

\graphicspath{{./Figures/}}
\usepackage[
bookmarks=true,
colorlinks=true,
linkcolor=blue,
urlcolor=blue,
citecolor=blue,
pdftex,
bookmarks=true,
linktocpage=true, % makes the page number as hyperlink in table of content
hyperindex=true
]{hyperref}

\usepackage{chemformula}
\usepackage{bibunits} % for generating separate bibilographies for main text and SI
\usepackage{physics}

\DeclareSIUnit \belm {Bm}
\DeclareSIUnit \dbm {dBm}
\newcommand{\epflAddress}{Institute of Physics, Swiss Federal Institute of Technology Lausanne (EPFL), CH-1015 Lausanne, Switzerland}
\newcommand{\cqsAddress}{Center of Quantum Science and Engineering (EPFL), CH-1015 Lausanne, Switzerland}
\newcommand{\purdueAddress}{OxideMEMS lab, Purdue University, West Lafayette, IN, USA}
\newcommand{\maxTotalEfficiency}{$1.6 \times 10^{-5}$}
\newcommand{\maxTotalEfficiencyDecibel}{\SI{-48}{\decibel}}
\newcommand{\maxOnChipEfficiency}{$7.9 \times 10^{-5}$}
\newcommand{\maxOnChipEfficiencyDecibel}{\SI{-41}{\decibel}}
\newcommand{\maxInternalEfficiency}{\SI{2e-3}{}}

\newcommand{\maxPumpPower}{\SI{21}{\deci\belm}}
\newcommand{\bandwidth}{\SI{25}{\mega\hertz}}

\newcommand*{\hham}{\hat{\mathcal{H}}} % Hamiltonian
\newcommand{\SIadj}[2]{\SI[number-unit-product={\text{-}}]{#1}{#2}} % connected with hyphen

\begin{document}

\title{Bidirectional microwave-optical transduction based on integration of high-overtone bulk acoustic resonators and photonic circuits}

\author{Terence Bl\'esin}
\affiliation{\epflAddress}
\affiliation{\cqsAddress}

\author{Wil Kao}
\affiliation{\epflAddress}
\affiliation{\cqsAddress}

\author{Anat Siddharth}
\affiliation{\epflAddress}
\affiliation{\cqsAddress}

\author{Rui N. Wang}
\affiliation{\epflAddress}
\affiliation{\cqsAddress}

\author{Alaina Attanasio}
\affiliation{\purdueAddress}

\author{Hao Tian}
\affiliation{\purdueAddress}

\author{Sunil A. Bhave}
\email[]{bhave@purdue.edu}
\affiliation{\purdueAddress}

\author{Tobias J. Kippenberg}
\email[]{tobias.kippenberg@epfl.ch}
\affiliation{\epflAddress}
\affiliation{\cqsAddress}

\begin{abstract}
	Coherent interconversion between microwave and optical frequencies can serve as both classical and quantum interfaces for computing, communication, and sensing. 
	Here, we present a compact microwave-optical transducer based on monolithic integration of piezoelectric actuators atop silicon nitride photonic circuits. 
	Such an actuator directly couples microwave signals to a high-overtone bulk acoustic resonator defined by the suspended silica cladding of the optical waveguide core,
	which leads to enhanced electromechanical and optomechanical couplings. 
	At room temperature, this triply resonant piezo-optomechanical transducer achieves an off-chip photon number conversion efficiency of \maxTotalEfficiency~over a bandwidth of \bandwidth~at an input pump power of \maxPumpPower. 
	The approach is scalable in manufacturing and, unlike existing electro-optic transducers, does not rely on superconducting resonators. 
	As the transduction process is bidirectional, we further demonstrate synthesis of microwave pulses from a purely optical input.
	Combined with the capability of leveraging multiple acoustic modes for transduction, the present platform offers prospects for building frequency-multiplexed qubit interconnects and for microwave photonics at large.
\end{abstract}

\maketitle

%%%%%%%%%%%%%%%%%%%%%%%%%%%%%%%%%%%%%%%%%%%%%%%%%%%%%%%%%%%%%%%%%%%%%%%%%%%%%%%%%%%%%%%%%%%%%%%%%%%%
%	Structure in journal-agnostic fashion: compatible with e.g. Optica, Nat Comm
%%%%%%%%%%%%%%%%%%%%%%%%%%%%%%%%%%%%%%%%%%%%%%%%%%%%%%%%%%%%%%%%%%%%%%%%%%%%%%%%%%%%%%%%%%%%%%%%%%%%
\begin{bibunit}[apsrev4-2] % biblography generated for main text defined within
\section{Introduction}
Modern data centers have seen rapidly increasing traffic that motivates an overhaul of existing network infrastructures.
As optical fibers support nearly lossless transport and high bandwidths, twisted-pair copper deployment has shrunk in favor of optical interconnects.
Energy-efficient optical transceivers \cite{leeEnergyEfficiencyOptical2012} and optical network architectures \cite{chengRecentAdvancesOptical2018} are being explored in parallel to accommodate emerging data- and resource-intensive applications.
In an analogous fashion, processing quantum information in superconducting circuits and networking via photonic interconnects has been envisioned as an effective strategy to address the scalability challenges in advancing quantum technologies \cite{ciracQuantumStateTransfer1997,duanLongdistanceQuantumCommunication2001,awschalomDevelopmentQuantumInterconnects2021}.
The scheme, featuring the full universal set of microwave quantum gates \cite{chowUniversalQuantumGate2012} with vanishing thermal occupancy and loss of the optical channels, 
%as well as the quantum optics toolbox available in the optical domain.
calls for the development of microwave-optical transducers to bridge the two energy scales that differ by more than four orders of magnitude.
Aside from facilitating the networking of remote quantum processors, these transducers may enable fully optical control and readout of microwave qubits \cite{lecocqControlReadoutSuperconducting2021,youssefiCryogenicElectroopticInterconnect2021}.
The subsequent replacement of coaxial lines bridging room temperature and cryogenic environments by optical fibers is expected to significantly ease the space and heat load constraints in dilution refrigerators, opening up a path toward upscaling processor units housed in a single fridge.

Efficient frequency conversion requires a nonlinear interaction stronger than the coupling to loss channels. 
The highest conversion efficiency to date has been achieved using an electro-optomechanical approach \cite{andrewsBidirectionalEfficientConversion2014}. 
There, a silicon nitride membrane interacts with an optical mode of a free-space Fabry-P\'erot cavity via radiation pressure and simultaneously serves as the top plate of a capacitor, parametrically coupling the mechanics to the microwave resonator.
By virtue of the high resonator quality factors and pump power handling capability, $47$\% of input photons can be interconverted between optical and microwave domains \cite{higginbothamHarnessingElectroopticCorrelations2018}, approaching the $50$\% efficiency required for attaining finite quantum capacity \cite{wolfQuantumCapacitiesBosonic2007}. 
This highly efficient transducer has enabled optical dispersive qubit readout with negligible excess backaction \cite{delaneySuperconductingqubitReadoutLowbackaction2022}.
The low-frequency (\SI{}{\mega\hertz}) mechanical intermediary nevertheless limits the transduction bandwidth and leads to appreciable added noise even at dilution refrigerator temperature.
To address these drawbacks, piezo-optomechanical transducers based on optomechanical crystals (OMC) have been developed \cite{bochmannNanomechanicalCouplingMicrowave2013,balramCoherentCouplingRadiofrequency2016, vainsencherBidirectionalConversionMicrowave2016,jiangLithiumNiobatePiezooptomechanical2019,forschMicrowavetoopticsConversionUsing2020,jiangEfficientBidirectionalPiezooptomechanical2020,mirhosseiniSuperconductingQubitOptical2020,honlMicrowavetoopticalConversionGallium2022,jiangOpticallyHeraldedMicrowave2023,weaverIntegratedMicrowavetoopticsInterface2022,meesalaNonclassicalMicrowaveopticalPhoton2023}, where a tightly confined high-frequency (\SI{}{\giga\hertz}) mechanical mode and a co-localized optical mode can interact at a vacuum optomechanical coupling rate $g_0 \sim 2\pi\times 500$~kHz. Other piezo-optomechanical transduction platforms have also been developed \cite{xiongCavityPiezooptomechanicsPiezoelectrically2013, hanCavityPiezomechanicsSuperconductingnanophotonic2020}. 
The trade-off lies instead in the sophistication required for microwave-phonon wave matching that significantly increases design complexity, as well as thermo-optic instability that constrains the intra-cavity photon number.
An on-chip efficiency of $5$\% has recently been reported on such a platform \cite{jiangOpticallyHeraldedMicrowave2023}.
However, low thermal conductance of these suspended quasi-one-dimensional structures hinders correlated microwave-optical photon pair generation at a practical rate. 
 
Cavity electro-optic modulators constitute a conceptually simpler approach where Pockels effect directly mediates microwave-optical interaction \cite{tsangCavityQuantumElectrooptics2010,tsangCavityQuantumElectrooptics2011,javerzac-galyOnchipMicrowavetoopticalQuantum2016}. 
On-chip realizations employing planar superconducting microwave resonators, benefited from the deep sub-wavelength mode volume of the vacuum electric field, to reach single-photon electro-optic coupling rates $g_0 \sim 2\pi\times 1$~kHz \cite{fanSuperconductingCavityElectrooptics2018,holzgrafeCavityElectroopticsThinfilm2020,mckennaCryogenicMicrowavetoopticalConversion2020,xuBidirectionalInterconversionMicrowave2021}.
However, the material science of $\chi^{(2)}$ crystals poses additional challenges.
Photorefractive effects---observed, for instance, in  \ch{LiNbO3} thin films \cite{liPhotonlevelTuningPhotonic2019}---hamper optical power handling, while piezoelectric loss and scattered optical photons degrade the quality factor of superconducting resonators \cite{dinizIntrinsicPhotonLoss2020,xuLightInducedDynamicFrequency2022}.
The difficulty of producing smooth sidewall surfaces in the workhorse Pockels material, \ch{LiNbO3}, through dry etching results in propagation losses an order of magnitude above the absorption limit \cite{zhuIntegratedPhotonicsThinfilm2021}.
As a result, the maximum on-chip efficiency achieved with integrated electro-optic transducers exceeds just $2$\% \cite{fanSuperconductingCavityElectrooptics2018}.
In the spirit of Ref.~\cite{andrewsBidirectionalEfficientConversion2014}, a bulk transducer comprising a mm-size mechanically polished \ch{LiNbO3} whispering gallery mode (WGM) resonator coupled to a 3D superconducting microwave cavity has proved competitive, trading $g_0$ for improved power handling and quality factors \cite{ruedaEfficientMicrowaveOptical2016,heaseBidirectionalElectroOpticWavelength2020}.
Taking one step further with pulsed optical pumping, the system has demonstrated not only a hallmark $14.4$\% total efficiency but also electro-optic dynamical backaction \cite{sahuQuantumenabledOperationMicrowaveoptical2022,qiuCoherentOpticalControl2023} and microwave-optical quadrature entanglement \cite{sahuEntanglingMicrowavesLight2023}. 

With the aforementioned design trade-offs in mind, we present a new piezo-optomechanical transducer based solely on wafer-scale, CMOS-compatible fabrication processes.
We utilize a low-loss silicon nitride (\ch{Si3N4}) photonic molecule, 
as well as multiple \SI{}{\giga\hertz} high-overtone bulk acoustic resonances (HBAR) parametrically coupled to the optical modes.
Endowed with power handling capabilities superior to existing integrated transducers, the device fits compactly within a \SI{100}{\micro\meter}-by-\SI{50}{\micro\meter} footprint, in contrast to state-of-the-art bulk designs \cite{higginbothamHarnessingElectroopticCorrelations2018,sahuQuantumenabledOperationMicrowaveoptical2022}.
In addition, the transduction HBAR modes are more readily coupled to the microwave signal, unlike OMC transducers.
We demonstrate bidirectional microwave-optical transduction with a bandwidth of \bandwidth~and total efficiency up to \maxTotalEfficiency~by pumping with \maxPumpPower~of off-chip optical power in continuous-wave (CW) operation.
The device is also characterized with a pulsed optical pump, which constitutes the first step toward optical control and readout of qubits, as well as heralded microwave-optical photon pair generation.
The simple design, ease of fabrication, robust operation, and compact form factor anticipate wide applicability in quantum technologies and microwave photonics at large.

\section{Results}
\subsection{Physics and design} 
\begin{figure*}[htb!]
	\centering
	\includegraphics[width=\linewidth]{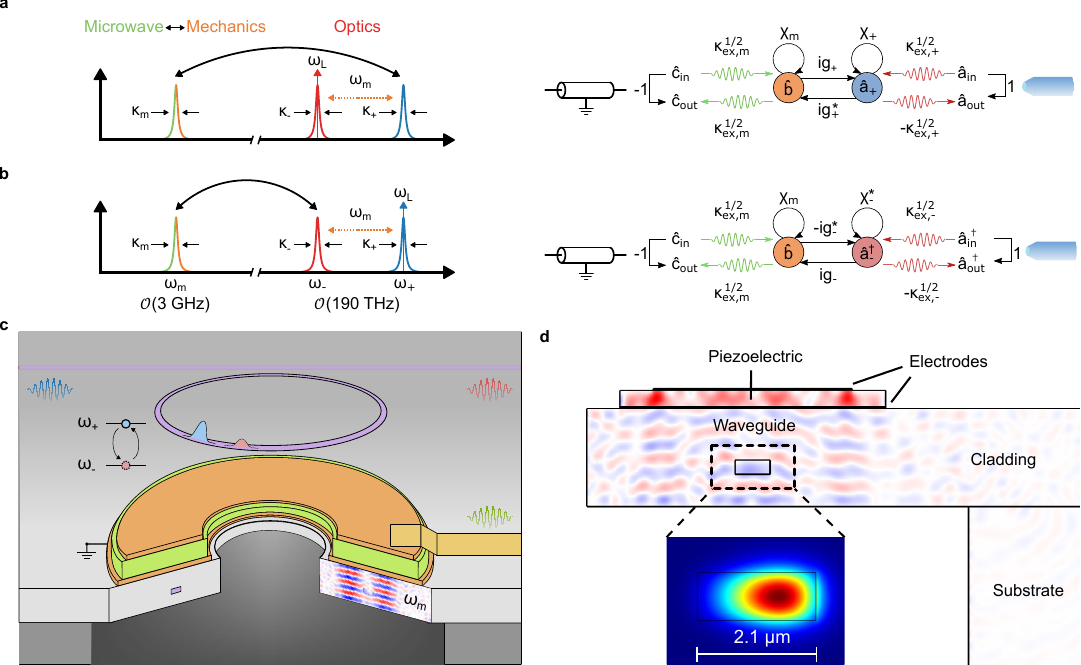}
	\caption{
		\textbf{Triply resonant piezo-optomechanical system.}
		\textbf{a,b}~Signal flow graphs illustrating the linearized equations of motion of the system, where the optical pump at frequency $\omega_L$ is set to the anti-Stokes and Stokes configurations, respectively.
		The optical and acoustic modes with susceptibilities $\chi$ are coupled with (multi-photon) rates $g_\pm = g_{0,\pm}\alpha$, where $\alpha$ is the pump optical field amplitude.
		Photons are emitted into (out of) the itinerant modes of input optical fiber $\hat{a}_\mathrm{out}$ ($\hat{a}_\mathrm{in}$) and microwave transmission line $\hat{c}_\mathrm{out}$ ($\hat{c}_\mathrm{in}$) at rates $\sqrt{\kappa_\mathrm{ex}}$.
		The acoustic and optical modes employed for transduction based on cavity-enhanced three-wave mixing are shown in frequency domain on the left.
		\textbf{c}~Experimental realization of the transducer (not to scale).
		A piezoelectric actuator (electrodes in orange and yellow; piezoelectric thin film in green) is integrated atop the photonic circuits (purple), allowing the acoustic modes supported by the suspended cladding (light grey) to be coupled to the microwave transmission line.
		The optical and microwave ports are labeled by colored wavelets.
		\textbf{d}~Stack composition and finite-element simulation of the transduction process.
		Both the mechanical stress pattern of the acoustic mode and the optical TE mode of the micro-ring are shown in the cross section.
	}
	\label{fig:fig_1}
\end{figure*}

Figure~\ref{fig:fig_1} delineates the principle of operation of the present transducer, while the theoretical formalism is detailed in Appendix~\ref{supp:theory}. 
The requisite nonlinear interaction for microwave-optical transduction is a parametric three-wave mixing process described by standard optomechanical Hamiltonian
\begin{equation*}\label{eq:hamiltonian}
	\hham_\mathrm{int} = - \hbar g_0\hat{a}^\dagger\hat{a}\left(\hat{b}+\hat{b}^\dagger\right).
\end{equation*}
Through the bilinear piezoelectric interaction, the acoustic resonance ($\hat{b}$) is coupled to the itinerant transmission line mode ($\hat{c}_\mathrm{in}$) of the same frequency $\omega_\mathrm{in} = \omega_m \approx \SI{3.5}{\giga\hertz}$. 
This frequency matches the detuning of the optical pump from the optical resonance ($\hat{a}$).
Specifically, an optical pump addressing the red (blue) side of the resonance induces an effective beam-splitter (two-mode-squeezing) interaction between $\hat{a}$ and $\hat{b}$, as illustrated by the signal flow graph in Fig.~\ref{fig:fig_1}a (\ref{fig:fig_1}b).
The primary figure of merit for a transducer is its efficiency, defined as the number of output photon number for each input photon.
An efficient device necessitates strong nonlinear interaction between the modes of interest, as well as ease of coupling to these modes internal to the device.
Our design addresses these two aspects.
First, the internal efficiency,
\begin{equation}\label{eq:internal_efficiency}
	\eta^\mathrm{int}_\pm = \frac{4C}{\left(1\pm C\right)^2},
\end{equation}
depends solely on the three-wave mixing cooperativity $C = 4g_0^2\bar{n}/\left(\kappa_o\kappa_m\right)$.
Here, $\bar{n}$, $\kappa_o$, and $\kappa_m$ denote the photon number in the optical cavity, and the total optical and acoustic linewidths of the transduction modes respectively.
The single photon optomechanical coupling rate $g_0 = - \left(\partial\omega_a/\partial x\right)x_\mathrm{ZPF}$, a product of the optical cavity frequency-pull parameter and zero-point displacement of the acoustic wave.
In Eq.~\ref{eq:internal_efficiency}, the plus and minus signs in the denominator correspond to the scenario where the pump is red- and blue-detuned (``anti-Stokes'' and ``Stokes'' processes), respectively.
Using a triply resonant configuration as in Refs.~\cite{fanSuperconductingCavityElectrooptics2018,holzgrafeCavityElectroopticsThinfilm2020,mckennaCryogenicMicrowavetoopticalConversion2020,xuBidirectionalInterconversionMicrowave2021} enhances the intracavity photon number $\bar{n}$ for a given on-chip input pump power $\eta^\mathrm{fiber-chip} P_\mathrm{in}$, where $\eta^\mathrm{fiber-chip}$ denotes the fiber-chip coupling efficiency and $P_\mathrm{in}$ the power in the optical fiber.
The corresponding photon flux is $\dot{n}_\mathrm{in} = \eta^\mathrm{fiber-chip} P_\mathrm{in}/\left(\hbar\omega_L\right)$.
The improved photon pumping efficiency can be explicitly seen by considering a ``hot'' cavity with coupling rate $\kappa_{\mathrm{ex},o}$ driven at a detuning $\Delta = \pm \omega_m$, yielding
\begin{equation*}
	\frac{\bar{n}}{\dot{n}_\mathrm{in}} = \frac{\kappa_{\mathrm{ex},a}}{\kappa_o^2/4+\Delta^2}.
\end{equation*}
Introducing an additional cavity mode centered at the pump frequency $\omega_L$ makes $\Delta = 0$, leading to an $\bar{n}$ enhanced by $\mathcal{O}\left(\omega_m^2/\kappa_o^2\right) = \mathcal{O}(10^3)$ and a significantly reduced laser power overhead required for a given $\eta^\mathrm{int}$.

The on-chip conversion efficiency $\eta^\mathrm{oc}$ is related to $\eta^\mathrm{int}$ by a proportionality constant
\begin{equation}\label{eq:extraction_efficiency}
	\eta^\mathrm{ext} = \frac{\kappa_{\mathrm{ex},o}}{\kappa_o}\frac{\kappa_{\mathrm{ex},m}}{\kappa_m}
\end{equation}
such that $\eta^\mathrm{oc} = \eta^\mathrm{ext}\eta^\mathrm{int}$. 
This extraction efficiency denotes the fraction by which optical and microwave photons are emitted out of the resonator modes with rates $\kappa_{\mathrm{ex},o}$ and $\kappa_{\mathrm{ex},m}$.
Our strategy of optimizing $\eta^\mathrm{ext}$ is most directly seen in the device implementation, illustrated in Fig.~\ref{fig:fig_1}d and proposed in Ref.~\cite{blesinQuantumCoherentMicrowaveoptical2021}.
The vertical stack comprises a piezoelectric actuator, a \ch{Si3N4} photonic waveguide layer, and a silica (\ch{SiO2}) cladding that also serves as the acoustic cavity \cite{tianHybridIntegratedPhotonics2020,tianMagneticfreeSiliconNitride2021}.
The actuator, deposited on a standard silicon wafer after fabrication of the optical circuitry using the photonic Damascene process, is composed of a $c$-cut aluminum nitride (\ch{AlN}) thin film sandwiched between top and bottom molybdenum (\ch{Mo}) electrodes on which the microwave signal is applied.
The alternating electric field launches longitudinal bulk acoustic waves (BAWs) into the cladding layer, which is suspended by etching away the silicon substrate to confine HBARs.
Since the \ch{AlN} is only located above the suspended cladding, the HBAR modes do not extend into the \ch{Si} substrate (Appendices~\ref{supp:process_flow} and \ref{supp:acoustics}).
These cavity-enhanced BAWs modify the waveguide refractive index through photoelastic and moving boundary effects, realizing the three-wave mixing interaction with rate $g_0$ \cite{blesinQuantumCoherentMicrowaveoptical2021}.
The stack composition is illustrated in Fig.~\ref{fig:fig_1}d.
Resonances that simultaneously satisfy the interfacial acoustic boundary conditions defined by the actuator and the whole stack naturally feature sizable $\kappa_{\mathrm{ex},m}$.
The simplicity of our design is in stark contrast to piezo-optomechanical transducers based on OMCs.
To facilitate efficient microwave extraction, design challenges need to be overcome to achieve strong hybridization between the piezo mode supported by the phonon waveguide and the transduction mode of the OMC cavity.

Central to both $\eta^\mathrm{int}$ and $\eta^\mathrm{ext}$ are the resonator intrinsic quality factors, as seen in Eqs.~\ref{eq:internal_efficiency} and~\ref{eq:extraction_efficiency}.
Unlike superconducting microwave resonators, HBARs based on amorphous \ch{SiO2} have been shown to exhibit sufficiently low (acoustic) loss at room temperature, relaxing the requirement for cryogenic operation \cite{tianHybridIntegratedPhotonics2020}.
A photonic molecule comprising a pair of evanescently coupled micro-rings can be hybridized into a symmetric mode ($\hat{a}_{-}$) and an antisymmetric mode ($\hat{a}_{+}$), which in turn make up the optical portion of the triple resonance system.
The waveguides are fabricated using the photonic Damascene process with wafer-scale yield \cite{pfeifferUltrasmoothSiliconNitride2018,liuHighyieldWaferscaleFabrication2021}.
The thick \ch{Si3N4} core reduces bending loss, allowing us to employ micro-rings with a radius (free spectral range) of \SI{22}{\micro\meter} (\SI{1}{\tera\hertz}).
As illustrated in Figs.~\ref{fig:fig_1}c and \ref{fig:fig_2}a, the micro-ring defines the dimensions of the piezoelectric actuator.
Reducing the micro-ring size therefore results in a reduced acoustic mode volume, which serves to increase $x_\mathrm{ZPF} \propto 1/\sqrt{V_m}$ and hence $g_0$.
 
\subsection{Device characterization}
We perform metrological characterization to affirm the integrity of the stack.
Shown in Fig.~\ref{fig:fig_2}a, the region of suspended cladding is first identified through optical micrography; the released substrate below leads to a contrast in the image.
To further access the layer structure, we use focused ion beam milling to create an opening on the device surface at the suspension site.
This opening provides sufficient clearance to directly image the cladding acoustic resonator, pictured in Fig.~\ref{fig:fig_2}b, confirming the removal of silicon.

We then characterize the transducer as an optoelectronic network with one microwave port, two optical ports, and one auxiliary electrical DC port.
Each optical port comprises an edge-coupled lensed fiber and on-chip bus waveguide terminated by 1D nanotapers designed for TE-polarized light, yielding a fiber-to-fiber coupling efficiency of $\eta^\mathrm{fiber-chip} = \SI{-8}{\decibel}$. 
For adjusting the frequency splitting between optical supermodes, a bias voltage is applied to the DC port to drive either the piezoelectric actuator \cite{liuMonolithicPiezoelectricControl2020} or an integrated thermo-optic heater placed in the vicinity of the micro-rings (Appendix~\ref{supp:heaters}).
The former affords no additional heat load and is hence cryogenic-compatible, whereas the latter provides an easy alternative for fast room-temperature characterization.
The avoided mode crossing characteristic of a photonic molecule is exemplified in Fig.~\ref{fig:fig_2}c.
We access the HBARs through the microwave port. 
Calibrated high-frequency probes are used to contact the top and bottom electrode pads of the piezoelectric actuator, effectively linking the acoustic resonator with a transmission line.
The reflection spectrum (Fig.~\ref{fig:fig_2}d) reveals predominantly one single series of HBARs with a free spectral range (FSR) of \SI{320}{\mega\hertz}, which corresponds to the acoustic length of the cladding.
The transduction acoustic mode at \SI{3.48}{GHz} exhibits a typical total linewidth of $\kappa_m/(2\pi) = \SI{13}{MHz}$ with a microwave extraction efficiency of $\kappa_{\mathrm{ex},m}/\kappa_m \approx 11\%$ (Supplementary Fig.~\ref{suppfig:mechanical_modes}a).
The microwave response is distinct from that of an identical stack composition with unreleased substrate. 
There, a periodic envelope corresponding to the cladding modes that are more strongly coupled to the microwave is superimposed over the full-stack HBAR response (Appendix~\ref{supp:acoustics} and Ref.~ \cite{tianHybridIntegratedPhotonics2020}).
The absence of these full-stack modes provide another piece of evidence of successful cladding suspension.
Finally, we study the device acousto-optic response in the triply resonant configuration, where the transducer effectively operates as a resonant single-sideband modulator.
The optical output containing both the pump at the symmetric-mode frequency and generated sideband at the antisymmetric-mode frequency are mixed by a photodetector.
Such a beat-note spectrum shown in Fig.~\ref{fig:fig_2}e displays response peaks aligned with the HBAR frequencies in Fig.~\ref{fig:fig_2}d, demonstrating three-wave mixing.

\subsection{Bidirectional microwave-optical transduction}
We measure the off-chip photon number transduction efficiency $\eta^\mathrm{tot}$ of the microwave-to-optical (up-conversion) and optical-to-microwave (down-conversion) processes as a function of off-chip CW input optical pump power $P_\mathrm{in}$.
Compared to the on-chip efficiency $\eta^\mathrm{oc}$, this efficiency also accounts for the loss channels of the microwave and optical ports.
First, we study the transducer in the triply resonant configuration.
For up-conversion, the optical sideband generated from the microwave input, detuned by $\omega_\mathrm{in}=\omega_m$ from the pump frequency, is measured via a self-calibrated heterodyne detection method.
The off-chip optical output containing both the pump ($\omega_L$) and the sideband ($\omega_s$) is combined with a local oscillator (LO; $\omega_\mathrm{LO}$) and detected.
By placing the LO at a frequency $\omega_\mathrm{LO} = \left(\omega_L+\omega_s\right)/2+\delta$ such that the photodetector response is constant over a frequency span of $\delta$, we determine the sideband power relative to the pump.
Picking off a fraction of the optical output then enables the determination of its absolute power and, by extension, that of the sideband with a power meter.
The measured optical sideband power is compared to that of the microwave input to yield $\eta^\mathrm{tot}$.
In the case of down-conversion, the optical input is generated by modulating the phase of the pump with an electro-optic modulator (EOM) driven by a microwave source of frequency $\omega_\mathrm{EOM} = \omega_m$.
Only one of the resulting sidebands is admitted into the photonic molecule and transduced, as the other is far off-resonance.
The converted microwave power is directly probed with an electrical spectrum analyzer (ESA).
Summarized in Figs.~\ref{fig:fig_3}a and c, bidirectional transduction processes corresponding to both the effective beam-splitter ($\omega_L=\omega_-$; anti-Stokes) and two-mode-squeezing ($\omega_L=\omega_+$; Stokes) interactions are investigated. Fitting the power dependence of the transduction efficiency yields $g_0 \approx 2\pi \times \SI{42}{\hertz}$ (Appendix~\ref{supp:parameter_characterization}).
We reach a maximal $\eta^\mathrm{tot}$ of \maxTotalEfficiencyDecibel~at \maxPumpPower~pump for each configuration.
Knowing the port losses, we estimate an on-chip efficiency $\eta^\mathrm{oc} = \maxOnChipEfficiencyDecibel$ (Appendix~\ref{supp:parameter_characterization}). 
Accounting for the extraction efficiency $\eta^\mathrm{ext}$, we further obtain an internal conversion efficiency $\eta^\mathrm{int} = \maxInternalEfficiency$ from Eq.~\ref{eq:extraction_efficiency}.

Furthermore, we deviate from triple resonance to map out the transduction bandwidth.
First, for a pump still resonant with one of the optical supermodes, the input microwave frequency $\omega_\mathrm{in}$ and input optical detuning (controlled by EOM drive frequency $\omega_\mathrm{EOM}$) are varied for up- and down-conversion, respectively.
Shown in Fig.~\ref{fig:fig_3}b and d, transduction leveraging the main transduction mode exhibits a full width at half maximum (FWHM) of \bandwidth.
The multimode nature of the transducer is manifested in Fig.~\ref{fig:fig_3}e, where the pump is in the beam-splitter configuration but slightly detuned from $\hat{a}_{-}$.
An additional transduction peak in $\eta^\mathrm{tot}$ with a FWHM of \SI{10}{\mega\hertz} is observed around $\omega_\mathrm{in} = 2\pi\times\SI{3.165}{\giga\hertz}$, which corresponds to another HBAR one FSR away from the main transduction mode $\omega_m = 2\pi\times\SI{3.480}{\giga\hertz}$.
The engineering degrees of freedom such as cladding and actuator thickness (HBAR FSR), and optical dispersion (supermode splitting as a function of optical wavelength) in the present system offer possibilities for frequency-multiplexed transduction.

\subsection{Pulsed transduction}
Quantum-enabled operation, where the added noise referred to the transducer input is less than one quanta, would require concurrently attaining high conversion efficiency and low output noise.
As the three-wave mixing interaction is parametric in nature, the associated cooperativity and hence efficiency can in principle be enhanced by boosting the optical pump power.
A potential trade-off is nevertheless the subsequently increased thermal noise.
While the high resonance frequency of the transduction HBAR already helps suppressing this noise to some extent, additionally employing a pulsed pump presents a number of utilities.
Reducing the integrated optical power serves as an effective measure to mitigate thermal load onto the cryostat as well as pump-induced noise while maintaining high peak power.
Quantum-enabled operation has in fact been achieved with this strategy by a bulk electro-optic transducer employing Watt-scale optical pumping \cite{sahuQuantumenabledOperationMicrowaveoptical2022}.
Gate-based superconducting quantum computers too function inherently in the pulsed regime.
As such, we characterize our transducer in both frequency and time domains with a pulsed optical pump to evaluate its compatibility with these cryogenic microwave circuits as well as its potential for quantum-enabled operation.

To study pulsed bidirectional transduction, we choose a pulse-on time $\tau_\mathrm{on} = \SI{1}{\micro\second}$ and a repetition rate $f_\mathrm{rep} = \SI{100}{\kilo\hertz}$.
We program the pulse sequence such that the increase in temperature, and hence acoustic mode thermal occupancy due to optical heating, is expected to be inconsequential (Appendix~\ref{supp:thermal_response}).
First, we measure the up-conversion efficiency, summarized in Fig.~\ref{fig:fig_4}a,  using the same heterodyne method as in the CW case (Fig.~\ref{fig:fig_3}).
The pulsed optical pump mediates transduction of a CW microwave input into a pulsed optical output.
This optical pulse comprising both the up-converted output and pump is then down-mixed by the LO, resulting in a microwave pulse whose frequency content can be probed via an ESA.
The measured efficiency exhibits reasonable agreement with CW data.
Additionally, the down-converted microwave pulse envelope in the time domain provides an independent measure of the transducer bandwidth---a key metric for qubit-photonic interconnects \cite{lecocqControlReadoutSuperconducting2021,youssefiCryogenicElectroopticInterconnect2021}.
With the pulsed pump set in the two-mode squeezing configuration, an optical input pulse is converted on-chip into a microwave pulse with a carrier frequency $\omega_\mathrm{EOM} = \omega_m$ through difference frequency generation.
The time-domain dynamics of the pulse envelope is captured through phase-sensitive demodulation using a digital lock-in amplifier.
As shown in Fig.~\ref{fig:fig_4}b, we observe that the pulse envelope is consistent with the step response of the lock-in integrator, which has a bandwidth of \SI{5}{\mega\hertz}, or an RC time constant $\tau_\mathrm{RC} = \SI{30}{\nano\second}$.
The lock-in bandwidth therefore sets a lower bound for the transducer bandwidth.

\section{Discussion}
Non-classically correlated microwave-optical photon pairs can be generated through spontaneous parametric down-conversion in our transducer.
These correlated photon pairs constitute a key ingredient toward entangling distant quantum processors via the Duan-Lukin-Cirac-Zoller protocol \cite{duanLongdistanceQuantumCommunication2001,krastanovOpticallyHeraldedEntanglement2021}.
The current on-chip pair generation rate is estimated to be \SI{1.5}{\kilo\hertz}, which is well above the thermal decoherence rate for \SI{3.5}{\giga\hertz} at \SI{10}{\milli\kelvin}, as explained in Appendix~\ref{supp:pair_rates}.
However, losses in the measurement setup specific to such an experiment, as well as the gating of the optical pump required to alleviate the effective heat load, are expected to limit further the final heralding rate.
Quasi-free-standing structures, such as 1D OMCs, do not readily thermalize and are thus more susceptible to heating effects.
Even though our transducer also utilizes a suspended acoustic resonator, it may be feasible to operate at a higher duty cycle than what is presented in Fig.~\ref{fig:fig_4}. 
It has been shown that a buffer gas environment can facilitate thermalization of a 1D OMC, at the cost of increased damping of the mechanical breathing mode \cite{shomroniOpticalBackactionevadingMeasurement2019}.
On the contrary, the HBAR mode of interest here should be relatively insensitive to such viscous damping, as the acoustic wave propagates predominantly along the longitudinal direction inside the cladding.
Therefore, a sample cell affixed to the mixing chamber flange and filled with buffer superfluid helium---an inviscid fluid that is both an excellent thermal conductor and electrical insulator---could potentially improve the heralding rate without hampering the conversion efficiency \cite{kashkanovaOptomechanicsSuperfluidHelium2017}.

The present design is also suited for tasks beyond quantum state transfer. 
The transducer can be a key component of photonic interconnects for superconducting qubits \cite{lecocqControlReadoutSuperconducting2021,youssefiCryogenicElectroopticInterconnect2021}, where laser lights routed through optical fibers are used to encode microwave signals directly inside the dilution refrigerator during qubit readout.
It may be challenging for traveling-wave EOMs to attain the requisite half-wave voltage to be competitive in noise performance against conventional all-electrical readout scheme utilizing high-electron-mobility transistor amplifiers \cite{youssefiCryogenicElectroopticInterconnect2021}. 
With a compact footprint of \SI{100}{\micro\meter}-by-\SI{50}{\micro\meter} and more than \SI{0.1}{\pico\watt} of microwave power generated on-chip, our triply resonant approach affords a path towards a scalable optically controlled cryogenic waveform generator for both qubit readout and control.
The multimode nature of the transducer is particularly suited for frequency-multiplexed dispersive readout.
Transduction leveraging multiple HBAR modes is already demonstrated in Fig.~\ref{fig:fig_3}e.
One can additionally utilize line-type instead of point-type dimer coupling for the photonic molecule, which gives rise to a dispersion in the supermode splitting \cite{kimDispersionEngineeringFrequency2017}.
In combination with the multitude of acoustic overtones, a single transducer can therefore support several spectrally distinguishable triply resonant systems that serve as qubit multiplex channels.
Finally, since the transducer requires no superconducting element to function, it may be pertinent for applications in classical microwave photonics.

Optimization of design and fabrication should engender further improvement on device performance.
The conversion efficiency, proportional to microwave extraction efficiency (Eq.~\ref{eq:extraction_efficiency}), can be improved by replacing \ch{AlN} with scandium-doped \ch{AlN} \cite{akiyamaEnhancementPiezoelectricResponse2009}, which features a piezoelectric coefficient $d_{33}$ about five times larger than \ch{AlN} \cite{akiyamaInfluenceGrowthTemperature2009,yanagitaniElectromechanicalCouplingGigahertz2014} while preserving CMOS compatibility.
\ch{AlScN} provides the additional benefit of increasing the strain induced in the cladding for a given voltage, resulting in a single photon coupling rate $g_0$ twice higher for high \ch{Sc}-doping concentrations.
Using thinner optical waveguides would reduce acoustic scattering losses, improving the mechanical quality factor as well as the microwave extraction efficiency.
There is likewise additional upside on the photonics (Appendix~\ref{supp:linewidth}).
Bending radiation loss can be significantly reduced by increasing the micro-ring radius $r$ or using a material with higher refractive index.
In particular, intrinsic quality factor exceeding $10^7$ has been demonstrated with an $r$ approximately ten times larger using the same fabrication process \cite{liuUltralowpowerChipbasedSoliton2018}.
%Higher index waveguides, such as \ch{Si}, would confine the mode more thighly while also enhancing the $g_0$ and allowing for thinner waveguides, hence improving the mechanical quality factor by mitigating scattering losses.
%It may be anticipated that the efficiency gain here may be chiefly compensated by the reduction in optomechanical coupling rate $g_0 \sim 1/\sqrt{r}$.
%Nevertheless, by thinning the waveguide to \SI{200}{\nano\meter} that approaches half of the acoustic wavelength, $g_0$ is expected to remain unchanged.
As in the thick-core waveguide fabricated using the photonic Damascene process, such a thin-core waveguide on a subtractive manufacturing platform is expected to exhibit negligible bending loss \cite{siddharthHertzlinewidthFrequencyagilePhotonic2023}.
On the other hand, a higher index from waveguides such as \ch{Si} directly increases $g_0$, while also ensuring tighter optical confinement, providing further enhancement in $g_0$ by reducing the acoustic mode volume if sharper microresonator bends are used.
%Etching the optical waveguides in a material with higher photoelastic coefficients or higher refractive index than \ch{Si3N4}, for example \ch{Si}, could easily provide another factor of two improvement in $g_0$, and
%better control on the position of the waveguide inside the cladding can equally improve the single photon coupling rate by another factor of two if designing for a specific mode.
Precise control over the waveguide position within the cladding can result in an additional twofold improvement in $g_0$.
Furthermore, utilizing higher-order HBAR within the typical frequency range of superconducting qubits could offer a similar enhancement, provided that the thickness of the piezoelectric layer is adjusted to shift the electromechanical coupling envelope accordingly.
Assuming critical coupling condition on the optical side (Appendix~\ref{supp:optimal_coupling}), we expect another 1.5 times gain in the optical extraction efficiency $\eta^\mathrm{tot}$.
Finally, optical insertion losses can be nearly eliminated by employing a spot-size converter to facilitate mode-matching with lensed fibers \cite{zhuUltrabroadbandHighCoupling2016,yaoBroadbandHighefficiencyTripletip2020,bhandariCompactBroadbandEdge2020}.
Combining these improvements, this transducer could achieve an internal efficiency of $100 \%$, limited only by the overcoupling of the waveguides and the fiber-chip insertion.

In conclusion, we have designed, fabricated, and characterized a compact piezo-optomechanical microwave-optical transducer that integrates wafer-scale, CMOS-compatible HBAR and \ch{Si3N4} photonics technologies.
Free of any superconducting components, this triply resonant transducer attains a bidirectional off-chip photon number conversion efficiency of \maxTotalEfficiency~(\maxOnChipEfficiency~on-chip, \maxInternalEfficiency~internal) and a bandwidth of \bandwidth~for an input pump power of \maxPumpPower~at room temperature. 
Proof-of-principle experiments show  that these performances remain unchanged for pulsed optical pumping (\SI{1}{\micro\second} pulse width at \SI{100}{\kilo\hertz} repetition rate).
Multimode transduction leveraging distinct HBAR modes have been demonstrated and more than \SI{0.1}{\pico\watt} of microwave power has been generated directly on-chip, suggesting prospects for designing frequency-multiplexed photonic interconnects for superconducting qubits.
Realistic design and fabrication improvements may allow access to experiments in the quantum regime.

\section*{Acknowledgments}
The authors would like to thank Amirali Arabmoheghi, Nils Johan Engelsen and Junyin Zhang for experimental assistance and fruitful discussions.
This work was supported by the Air Force Office of Scientific Research under award no. FA8655-20-1-7009, Swiss National Science Foundation under grant agreement no. 204927, and European Research Council (ERC) under the EU H2020 research and innovation programme, grant agreement no. 835329 (ExCOM-cCEO).
This work was further supported by United States Air Force Research Laboratory Award no. FA8750-21-2-0500, as well as United States National Science Foundation’s International Collaboration Supplements in Quantum Information Science and Engineering Research for RAISE-TAQS Award no. 18-39164.
A.S. acknowledges support from the European Space Technology Centre with ESA Contract No. 4000135357/21/NL/GLC/my.
Samples were fabricated in the Center of MicroNanoTechnology (CMi) at EPFL and Birck Nanotechnology Center at Purdue University.

\section*{Author contributions}
T.B. designed the device. R.N.W, A.A. and H.T. fabricated the device. T.B., A.S. and W.K. conducted the experiments. T.B. and W.K. wrote the paper with input from all authors. T.J.K. and S.A.B. supervised the project.

\putbib[references]
\end{bibunit}

\begin{figure*}[htb!]
	\centering
	\includegraphics[width=\linewidth]{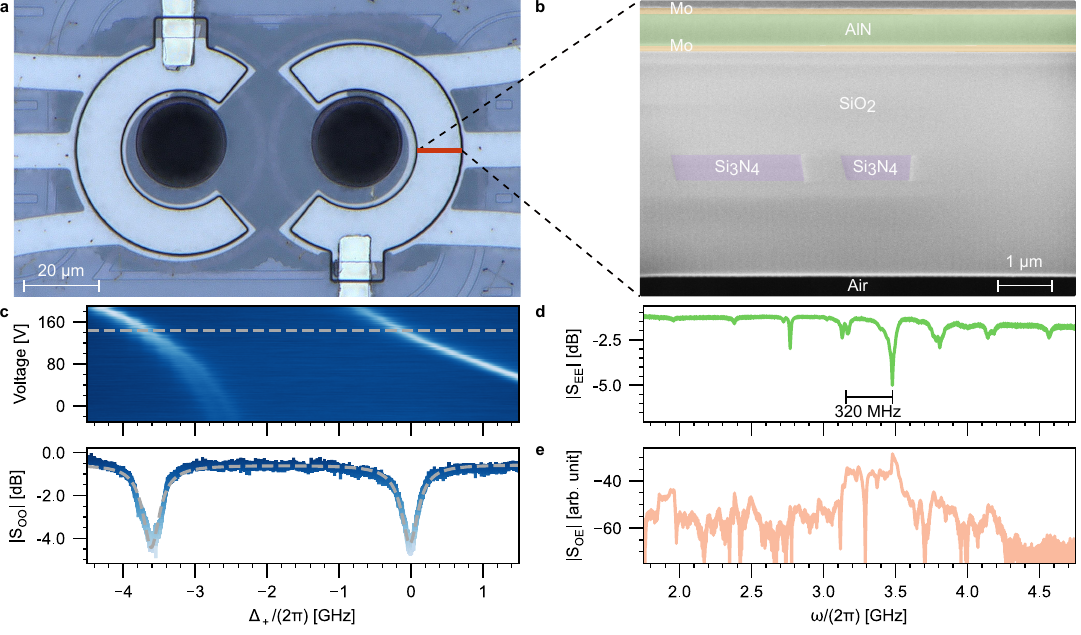}
	\caption{
		\textbf{Device metrological and network characterization.}
		\textbf{a}~Optical micrograph of the transducer. 
		The shaded region originates from the released silicon substrate.
		An additional actuator has been placed on the second ring cavity for DC stress-optic tuning of the optical resonances.
		\textbf{b}~False-color cross section of the device imaged by focused ion beam scanning electron microscopy showing the top and bottom electrodes (\ch{Mo}; yellow), piezoelectric layer (\ch{AlN}; green), suspended cladding (\ch{SiO2}; gray), and optical waveguides (\ch{Si3N4}; purple). 
		\textbf{c}~Optical transmission spectra near \SI{1550}{\nano\meter} for \SI{1}{\milli\watt} optical probe as a function of applied DC piezoelectric control voltage and detuning from the antisymmetric supermode.
		The transmission color map follows the representative spectrum in the bottom panel.
		The dashed line denotes the best-fit transmission from coupled mode theory, correpsonding to $\kappa_{r} = 2\pi \times \SI{154}{\mega\hertz}$ and $\kappa_{l} = 2\pi \times  \SI{190}{\mega\hertz}$ with $\kappa_{\mathrm{ex}, r} < 2\pi \times  \SI{10}{\mega\hertz}$ and $\kappa_{\mathrm{ex}, l} = 2\pi \times  \SI{120}{\mega\hertz}$ in the bare mode picture, and $\kappa_- = 2\pi \times \SI{166}{\mega\hertz}$ and $\kappa_+ = 2\pi \times  \SI{179}{\mega\hertz}$ with $\kappa_{\mathrm{ex}, -} = \kappa_{\mathrm{ex}, +} = 2\pi \times  \SI{60}{\mega\hertz}$ in the hybridized mode picture (Appendix~\ref{supp:hybrid}).
		\textbf{d}~Microwave reflection spectrum with a probe power of \SI{-10}{\deci\belm} indicating locations of the HBARs. 
		\textbf{e}~Acousto-optic response spectrum with an off-chip optical pump power $P_\mathrm{in}=\SI{20}{\deci\belm}$ centered at the symmetric supermode.
	}
	\label{fig:fig_2}
\end{figure*}

\begin{figure*}[htb!]
	\centering
	\includegraphics[width=\linewidth]{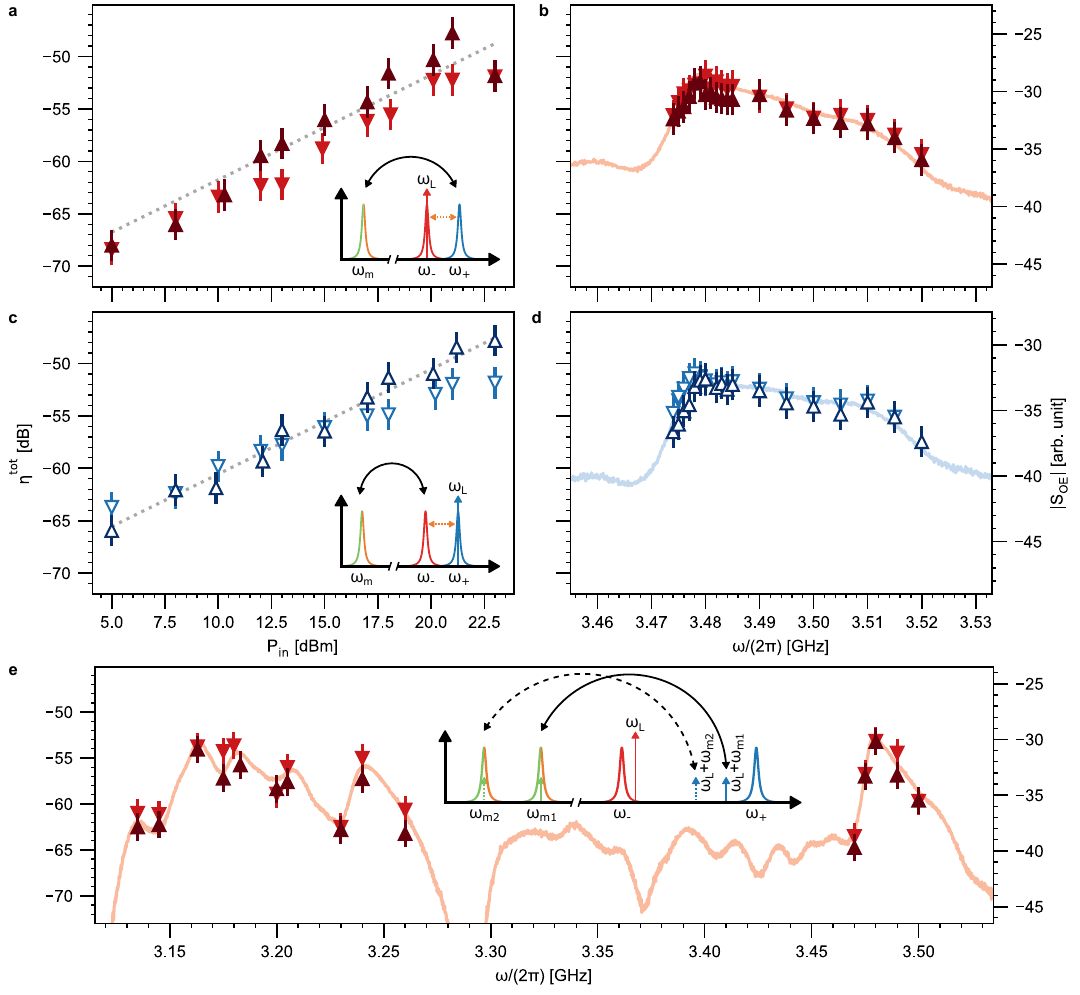}
	\caption{
		\textbf{Bidirectional, multimode transduction with a continuous-wave pump.}
		\textbf{a, b} Anti-Stokes transduction. 
		\textbf{c, d} Stokes transduction.
		Off-chip photon number transduction efficiencies as a function of off-chip input pump power in the triply resonant configuration are shown in the left panels. 
		Off-chip photon number transduction efficiencies as a function of input microwave frequency and input optical detuning are shown in the right panels, which depict the transduction bandwidth for an input optical pump power of \SI{20}{\deci\belm}.
		The up-conversion data were recorded with varying input microwave powers (\SI{-20}{\deci\belm}, \SI{-10}{\deci\belm} and \SI{0}{\deci\belm}) in randomized order. 
		An optical input 20 dB smaller than the optical pump was employed to measure down-conversion data.
		\textbf{e} Bidirectional transduction with a detuned pump leveraging multiple HBAR modes, as a function of input microwave frequency and input optical detuning for an input optical pump power of \SI{20}{\deci\belm}. 
		Up- and down-conversion efficiencies are denoted by triangles and inverted triangles, respectively, with the acoustic-optic response spectrum corresponding to a slight detuning superimposed.
		The markers are empty for Stokes processes and filled for anti-Stokes processes.
		The dotted lines represent the best-fit efficiency in the low-cooperativity regime where $\eta^\mathrm{tot} \propto C \propto P_\mathrm{in}$.
		The insets illustrate the respective configurations for the pump and input signal.
		\textbf{a} and \textbf{c} share the same power axis. \textbf{b} and \textbf{d} share the same frequency axis. All panels share the same off-chip photon number transduction efficiency axis.
	}
	\label{fig:fig_3}
\end{figure*}

\begin{figure*}[htb!]
	\centering
	\includegraphics[width=\linewidth]{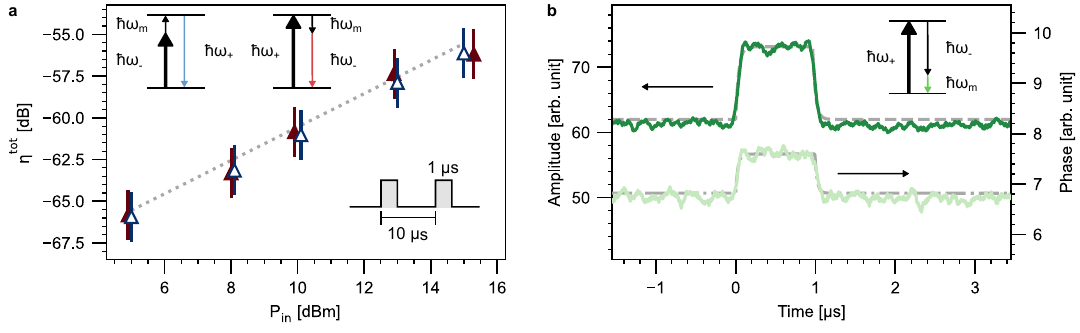}
	\caption{
		\textbf{Pulsed transduction.}
		\textbf{a}~Pulsed up-conversion as a function of off-chip input pump power for a pulse-on time of \SI{1}{\micro\second} and repetition rate of \SI{100}{\kilo\hertz}.
		The pump is in the Stokes (empty triangles) and anti-Stokes (filled triangles) configurations.
		The best-fit efficiency (dotted line) for CW operation from Fig.~\ref{fig:fig_3}c is shown for comparison.
		The microwave input powers have been randomized in the same way as for Fig.~\ref{fig:fig_3}.
		\textbf{b}~Down-converted microwave pulse in the time domain with a \SIadj{15}{\dbm} pump set to the Stokes configuration, detected as demodulated amplitude and phase averaged over 6000 shots.
		The dashed and dash-dotted lines are fits to the RC step response function that physically corresponds to the lock-in integrator, which has a bandwidth (RC time constant $\tau_\mathrm{RC}$) of \SI{5}{\mega\hertz} (\SI{30}{\nano\second}).
		The amplitude and phase response yields a best-fit $\tau_\mathrm{RC} = \SI{35\pm 4}{\nano\second}$ and $\SI{27\pm 3}{\nano\second}$, respectively. 
		These values are consistent with the lock-in bandwidth.
		The insets illustrate the respective configurations for the optical pump (thick black), input signal (black) and measured output (colored).
		The pulse sequence is shown in time domain.
	}
	\label{fig:fig_4}
\end{figure*}
\clearpage

%----------------------------------
%          Appendices
%----------------------------------
\setcounter{figure}{0}  
\renewcommand{\figurename}{Supplementary~Fig.} % without babel package

\begin{bibunit}[apsrev4-2] % begin section biblography section for the SI
\appendix
\onecolumngrid

\section{Theory}\label{supp:theory}

\subsection{Coupled mode theory for optical modes hybridization}\label{supp:hybrid}
Using a pair of optical resonances increases the intracavity photon number for a given input optical power, resulting in an enhanced optomechanical interaction.
Such an optical doublet is implemented here by evanescently coupling two micro-rings, forming a photonic molecule.
Consider two bare optical modes with resonance frequencies $\omega_l$ and $\omega_r$ that correspond to the left and right micro-rings, respectively.
The two modes couple with a rate $J$, which yields the Hamiltonian \cite{hausCoupledmodeTheory1991}
\begin{equation}
	\hham = \hbar \omega_l \hat{a}_l^{\dagger}\hat{a}_l + \hbar \omega_r \hat{a}_r^{\dagger}\hat{a}_r - \hbar J \left( \hat{a}_l \hat{a}_r^{\dagger} + \hat{a}_l^{\dagger} \hat{a}_r \right)
\end{equation}
and the equations of motion
\begin{equation}
		\begin{aligned}
			\frac{d}{dt} \hat{a}_l(t) &= \left(-i\omega_l - \frac{\kappa_l}{2}\right)\hat{a}_l(t) + i J \hat{a}_r(t), \\
			\frac{d}{dt} \hat{a}_r(t) &= \left(-i\omega_r - \frac{\kappa_r}{2}\right)\hat{a}_r(t) + i J \hat{a}_l(t). \\
		\end{aligned}
\end{equation}
With Laplace transform, they can be rewritten in frequency domain as
\begin{equation}
	\begin{pmatrix}
		\chi_l^{-1}[s] & -iJ \\
		-iJ & \chi_r^{-1}[s] \\
	\end{pmatrix}
	\begin{pmatrix}
		\hat{a}_l[s] \\
		\hat{a}_r[s]
	\end{pmatrix}
	=
	\begin{pmatrix}
		\hat{a}_l(0) \\
		\hat{a}_r(0)
	\end{pmatrix},
\end{equation}
where $\chi_o[s] = \left(s + i\omega_o + \kappa_o/2\right)^{-1}$, $o = l, r$ denote the susceptibilities.
Diagonalizing the matrix yields the eigenvalues
\begin{align}
	\lambda_{\pm} 
	&= \frac{\chi_l^{-1} + \chi_r^{-1}}{2} \pm \frac{1}{2} \sqrt{\left( \chi_l^{-1} - \chi_r^{-1} \right)^2 - 4 J^2}\nonumber\\
	&= \left( s + i \bar{\omega} + \bar{\kappa}/2 \right) \pm \frac{1}{2} \sqrt{\frac{\mu^2}{4} -\delta^2 + i \mu\delta - 4 J^2}.
\end{align}
Here we define the average optical frequency $\bar{\omega} = \left(\omega_l + \omega_r\right)/2$, average linewidth $\bar{\kappa} = \left(\kappa_l + \kappa_r\right)/2$, relative detuning of the micro-rings $\delta = \omega_l-\omega_r$, and linewidth difference $\mu = \kappa_l-\kappa_r$.
The difference in resonance frequency and linewidth of the two supermodes can be computed as
\begin{align}
	\Delta\omega &= \frac{1}{\sqrt{2}}\sqrt{-\frac{\mu^2}{4} + \delta^2 + 4 J^2 +\sqrt{ \left[ \left( 2J - \frac{\mu}{2} \right)^2 + \delta^2 \right] \left[ \left( 2J + \frac{\mu}{2} \right)^2 + \delta^2 \right] }},\\
	\Delta\kappa &= \sqrt{2}\sqrt{\frac{\mu^2}{4} -\delta^2 - 4 J^2 +\sqrt{ \left[ \left( 2J - \frac{\mu}{2} \right)^2 + \delta^2 \right] \left[ \left( 2J + \frac{\mu}{2} \right)^2 + \delta^2 \right] }},
\end{align}
where $\lambda_+ - \lambda_- = \Delta\kappa/2 + i\Delta\omega$. 
In the strong coupling regime where $2J\gg \mu/2$, we simply have $\Delta\omega \approx \sqrt{ 4 J^2 + \delta^2 }$ and $\Delta\kappa \approx \mu \sqrt{ 1 - 4J^2/(4 J^2 + \delta^2) }$.
The two supermodes are thus centered at $\bar{\omega}$ with a splitting of $\Delta\omega$, resulting in susceptibilites
\begin{equation}
	\chi_\pm[s] = \left[s + i \left( \bar{\omega} \pm \frac{\Delta\omega}{2} \right) + \frac{1}{2} \left(\bar{\kappa} \pm \frac{\Delta\kappa}{2}\right) \right]^{-1}.
\end{equation}
Solving for the eigenvectors gives the participation ratios of the bare cavity modes in the supermodes
\begin{equation}
	\begin{aligned}
		u_\pm &= \frac{2 J}{\sqrt{4 J^2 + |\mu \mp \Delta\kappa/2 + i\left( \delta \mp \Delta\omega \right)|^2}} \exp\left(i \phi_\pm\right), \\
		v_\pm &= \frac{|\mu \mp \Delta\kappa/2 + i\left( \delta \mp \Delta\omega \right)|}{\sqrt{4 J^2 + |\mu \mp \Delta\kappa/2 + i\left( \delta \mp \Delta\omega \right)|^2}} \exp\left[i\phi_\pm + i\arg{\frac{\mu \mp \Delta\kappa/2 + i\left(\delta\mp\Delta\omega\right)}{2iJ}}\right],
	\end{aligned}
\end{equation}
where $|u_\pm|^2 + |v_\pm|^2 = 1$ and $\phi_\pm$ is a global phase factor.
In particular, the supermode $\hat{a}_- = u_- \hat{a}_l + v_- \hat{a}_r$ has a resonance frequency $\omega_- = \bar{\omega} - \Delta\omega/2$ and linewidth $\kappa_- = \bar{\kappa} - \Delta\kappa/2$. 
It is referred to as the symmetric supermode.
The rationale behind the nomenclature can be seen by considering the limit of zero detuning $\delta=0$, where we simply have $\hat{a}_- = \left( \hat{a}_l + \hat{a}_r \right)/\sqrt{2}$ with $\omega_- = \bar{\omega} - J$ and $\kappa_- = \bar{\kappa}$.
On the other hand, the antisymmetric supermode $\hat{a}_+ = u_+ \hat{a}_l + v_+ \hat{a}_r$ has a resonance frequency $\omega_+ = \bar{\omega} + \Delta\omega/2$ and linewidth $\kappa_+ = \bar{\kappa} + \Delta\kappa/2$. 
When $\delta = 0$, this reduces to $\hat{a}_+ = \left( \hat{a}_l - \hat{a}_r \right)/\sqrt{2}$ with $\omega_+ = \bar{\omega} + J$ and $\kappa_+ = \bar{\kappa}$.

\subsection{Transducer Hamiltonian and input-output relations}
\label{supp:interaction_hamiltonian}
In terms of bare micro-ring modes, the transducer Hamiltonian is given by
\begin{equation}
	\hham = \hbar \omega_m \hat{b}^{\dagger} \hat{b} + \hbar \omega_l \hat{a}_l^{\dagger} \hat{a}_l + \hbar \omega_r \hat{a}_r^{\dagger} \hat{a}_r - \hbar J \left( \hat{a}_l^{\dagger} \hat{a}_r + \hat{a}_r^{\dagger} \hat{a}_l \right) 
	- \hbar g_0 \hat{a}_l^{\dagger} \hat{a}_l \left( \hat{b} + \hat{b}^{\dagger} \right)
	+ \hham_\mathrm{drive}.
\end{equation}
Note that the piezo-coupled acoustic mode interacts with only one of the micro-rings.
To compute quantities directly related to the experiments, we rewrite the Hamiltonian in terms of the supermodes (Appendix~\ref{supp:hybrid}) as
\begin{equation}
	\hham = \hbar \omega_m \hat{b}^{\dagger} \hat{b} + \hbar \omega_- \hat{a}_-^{\dagger} \hat{a}_- + \hbar \omega_+ \hat{a}_+^{\dagger} \hat{a}_+
	- \hbar g_0 \left( |x|^2 \hat{a}_-^{\dagger} \hat{a}_- + |y|^2 \hat{a}_+^{\dagger} \hat{a}_+ + xy^* \hat{a}_+^{\dagger} \hat{a}_- + x^*y \hat{a}_-^{\dagger} \hat{a}_+ \right) \left( \hat{b} + \hat{b}^{\dagger} \right)
	+ \hham_\mathrm{drive}
\end{equation}
such that $\hat{a}_l = x \hat{a}_- + y \hat{a}_+$.
The driving term is given by
\begin{equation}
	\begin{aligned}
		\hham_\mathrm{drive} =\ &i \hbar \sqrt{\kappa_\mathrm{ex, -}} \left( \hat{a}_\mathrm{in} \hat{a}_-^{\dagger} e^{-i\omega_L t} - \hat{a}_\mathrm{in}^{\dagger} \hat{a}_-e^{i\omega_L t} \right) 
		+ i \hbar \sqrt{\kappa_\mathrm{ex, +}} \left( \hat{a}_\mathrm{in} \hat{a}_+^{\dagger} e^{-i\omega_L t} - \hat{a}_\mathrm{in}^{\dagger} \hat{a}_+ e^{i\omega_L t} \right) \\
		&+\ i \hbar \sqrt{\kappa_\mathrm{ex, m}} \left( \hat{c}_\mathrm{in} \hat{b}^{\dagger} e^{i\omega_m t} - \hat{c}_\mathrm{in}^{\dagger} \hat{b} e^{-i\omega_m t} \right)
	\end{aligned}
\end{equation}
for input optical and microwave fields $\hat{a}_\mathrm{in}$ and $\hat{c}_\mathrm{in}$.
By going into the frames rotating at the laser drive frequency $\omega_L$, we obtain
\begin{equation}
	\begin{aligned}
		\hham =\ &\hbar \omega_m \hat{b}^{\dagger} \hat{b} - \hbar \Delta_- \hat{a}_-^{\dagger} \hat{a}_- - \hbar \Delta_+ \hat{a}_+^{\dagger} \hat{a}_+ \\
		& - \hbar g_0 \left( |x|^2 \hat{a}_-^{\dagger} \hat{a}_- + |y|^2 \hat{a}_+^{\dagger} \hat{a}_+ \right) \left( \hat{b} + \hat{b}^{\dagger} \right)
		- \hbar g_0 \left( xy^* \hat{a}_+^{\dagger} \hat{a}_- + x^*y \hat{a}_-^{\dagger} \hat{a}_+ \right) \left( \hat{b} + \hat{b}^{\dagger} \right) \\
		&+ i \hbar \sqrt{\kappa_\mathrm{ex, -}} \left( \hat{a}_\mathrm{in} \hat{a}_-^{\dagger} - \hat{a}_\mathrm{in}^{\dagger} \hat{a}_- \right)
		+ i \hbar \sqrt{\kappa_\mathrm{ex, +}} \left( \hat{a}_\mathrm{in} \hat{a}_+^{\dagger} - \hat{a}_\mathrm{in}^{\dagger} \hat{a}_+ \right)
		+ i \hbar \sqrt{\kappa_\mathrm{ex, m}} \left( \hat{c}_\mathrm{in} \hat{b}^{\dagger} e^{i\omega_m t} - \hat{c}_\mathrm{in}^{\dagger} \hat{b} e^{-i\omega_m t} \right),
	\end{aligned}
\end{equation}
where $\Delta_\pm = \omega_L - \omega_\pm$.
Considering small deviations of the supermodes around their steady-state amplitudes $\alpha_\pm \approx \left[\sqrt{\kappa_{\mathrm{ex},\pm}}/(\kappa_\pm/2-i\Delta_\pm)\right] \sqrt{\dot{n}_\mathrm{in}^\pm} e^{i\phi_\pm}$ and letting the input field operators denote just the Langevin force, the Hamiltonian
\begin{equation}
	\begin{aligned}
		\hham =\ &\hbar \omega_m \hat{b}^{\dagger} \hat{b} - \hbar \Delta_- \hat{a}_-^{\dagger} \hat{a}_- - \hbar \Delta_+ \hat{a}_+^{\dagger} \hat{a}_+ \\
		& - \hbar g_0 |x|^2 \left( |\alpha_-|^2 + \alpha_-^* \hat{a}_- + \hat{a}_-^{\dagger} \alpha_- + \hat{a}_-^{\dagger} \hat{a}_- \right) \left( \hat{b} + \hat{b}^{\dagger} \right) \\
		& - \hbar g_0 |y|^2 \left( |\alpha_+|^2 + \alpha_+^* \hat{a}_+ + \hat{a}_+^{\dagger} \alpha_+ + \hat{a}_+^{\dagger} \hat{a}_+ \right) \left( \hat{b} + \hat{b}^{\dagger} \right) \\
		& - \hbar g_0 xy^* \left( \alpha_+^* \alpha_- + \alpha_+^* \hat{a}_- + \hat{a}_+^{\dagger} \alpha_- + \hat{a}_+^{\dagger} \hat{a}_- \right) \left( \hat{b} + \hat{b}^{\dagger} \right) \\
		& - \hbar g_0 x^*y \left( \alpha_-^* \alpha_+ + \alpha_-^* \hat{a}_+ + \hat{a}_-^{\dagger} \alpha_+ + \hat{a}_-^{\dagger} \hat{a}_+ \right) \left( \hat{b} + \hat{b}^{\dagger} \right) \\
		&+ i \hbar \sqrt{\kappa_\mathrm{ex, -}} \left( \hat{a}_\mathrm{in} \hat{a}_-^{\dagger} - \hat{a}_\mathrm{in}^{\dagger} \hat{a}_- \right)
		+ i \hbar \sqrt{\kappa_\mathrm{ex, +}} \left( \hat{a}_\mathrm{in} \hat{a}_+^{\dagger} - \hat{a}_\mathrm{in}^{\dagger} \hat{a}_+ \right) \\
		&+ i \hbar \sqrt{\kappa_\mathrm{ex, m}} \left( \hat{c}_\mathrm{in} \hat{b}^{\dagger} e^{i\omega_m t} - \hat{c}_\mathrm{in}^{\dagger} \hat{b} e^{-i\omega_m t} \right)
	\end{aligned}
\end{equation}
can be further simplified.
In the large intracavity photon number limit, the $\mathcal{O}\left(\hat{a}^2\right)$ terms are ommited as their contribution to the system dynamics is hidden by the $\mathcal{O}\left(\hat{a}\right)$ terms.
The $\mathcal{O}\left(\alpha^2\right)$ constant terms are removed by shifting the position origin by the steady-state displacement
\begin{equation}
	x_{ss} = \frac{2 g_0}{\omega_m}\left( |x|^2 |\alpha_-|^2 + |y|^2 |\alpha_+|^2 + xy^* \alpha_+^* \alpha_- + x^*y \alpha_-^* \alpha_+ \right) x_\mathrm{ZPF}
\end{equation}
using the translation operator $\hat{T}_{\hat{x}}(x_{ss}) = \exp\left( x_{ss} \hat{p} / (i\hbar) \right)$, $\hat{p}$ being the momentum operator associated with $\hat{x} = x_\mathrm{ZPF} \left( \hat{b} + \hat{b}^{\dagger} \right)$.
The resulting effective Hamiltonian is then given by
\begin{equation}
	\begin{aligned}
		\hham =\ &\hbar \omega_m \hat{b}^{\dagger} \hat{b} - \hbar \Delta_- \hat{a}_-^{\dagger} \hat{a}_- - \hbar \Delta_+ \hat{a}_+^{\dagger} \hat{a}_+ \\
		& - \hbar g_0 \left( |x|^2 \alpha_-^* + xy^* \alpha_+^* \right) \hat{a}_- \left( \hat{b} + \hat{b}^{\dagger} \right) 
		- \hbar g_0 \left( |x|^2 \alpha_- + x^*y \alpha_+ \right) \hat{a}_-^{\dagger} \left( \hat{b} + \hat{b}^{\dagger} \right) \\
		& - \hbar g_0 \left( |y|^2 \alpha_+^* + x^*y \alpha_-^* \right) \hat{a}_+ \left( \hat{b} + \hat{b}^{\dagger} \right) 
		- \hbar g_0 \left( |y|^2 \alpha_+ + xy^* \alpha_- \right) \hat{a}_+^{\dagger} \left( \hat{b} + \hat{b}^{\dagger} \right) \\
		&+ i \hbar \sqrt{\kappa_\mathrm{ex, -}} \left( \hat{a}_\mathrm{in} \hat{a}_-^{\dagger} - \hat{a}_\mathrm{in}^{\dagger} \hat{a}_- \right)
		+ i \hbar \sqrt{\kappa_\mathrm{ex, +}} \left( \hat{a}_\mathrm{in} \hat{a}_+^{\dagger} - \hat{a}_\mathrm{in}^{\dagger} \hat{a}_+ \right) 
		+ i \hbar \sqrt{\kappa_\mathrm{ex, m}} \left( \hat{c}_\mathrm{in} \hat{b}^{\dagger} e^{i\omega_m t} - \hat{c}_\mathrm{in}^{\dagger} \hat{b} e^{-i\omega_m t} \right).
	\end{aligned}
\end{equation}
Finally, by defining effective optomechanical coupling rates 
\begin{equation}\label{eq:supermode_coupling}
	\begin{aligned}
		g_- &= g_0 \left( |x|^2 \alpha_- + x^*y \alpha_+ \right),\\
		g_+ &= g_0 \left( |y|^2 \alpha_+ + xy^* \alpha_- \right),
	\end{aligned}
\end{equation}
and going into a rotating frame defined by the operator
\begin{equation}
	\hat{U} = 
	\exp\left( \frac{1}{i\hbar} \hbar \Delta_- \hat{a}_-^{\dagger} \hat{a}_- t \right)
	\exp\left( \frac{1}{i\hbar} \hbar \Delta_+ \hat{a}_+^{\dagger} \hat{a}_+ t \right)
	\exp\left( \frac{-1}{i\hbar} \hbar \omega_m \hat{b}^{\dagger} \hat{b} t \right),
\end{equation}
we have
\begin{equation}
	\begin{aligned}
		\hham =\ & - \hbar \left( g_-^* \hat{a}_- e^{i \Delta_- t} + g_- \hat{a}_-^{\dagger} e^{-i \Delta_- t} \right) \left( \hat{b} e^{-i \omega_m t} + \hat{b}^{\dagger} e^{i \omega_m t} \right) \\
		 &- \hbar \left( g_+^* \hat{a}_+ e^{i \Delta_+ t} + g_+ \hat{a}_+^{\dagger} e^{-i \Delta_+ t} \right) \left( \hat{b} e^{-i \omega_m t} + \hat{b}^{\dagger} e^{i \omega_m t} \right) \\
		&+ i \hbar \sqrt{\kappa_\mathrm{ex, -}} \left( \hat{a}_\mathrm{in} \hat{a}_-^{\dagger} e^{-i\Delta_- t} - \hat{a}_\mathrm{in}^{\dagger} \hat{a}_- e^{i\Delta_- t} \right)
		+ i \hbar \sqrt{\kappa_\mathrm{ex, +}} \left( \hat{a}_\mathrm{in} \hat{a}_+^{\dagger} e^{-i\Delta_+ t} - \hat{a}_\mathrm{in}^{\dagger} \hat{a}_+ e^{i\Delta_+ t} \right) \\
		&+ i \hbar \sqrt{\kappa_\mathrm{ex, m}} \left( \hat{c}_\mathrm{in} \hat{b}^{\dagger} - \hat{c}_\mathrm{in}^{\dagger} \hat{b} \right)
	\end{aligned}
\end{equation}
in the interaction picture.
The time evolution of the bosonic modes can be computed using the Heisenberg-Langevin equations given by
\begin{equation}
	\begin{aligned}
		\frac{d}{dt} \hat{a}_-(t) = &\frac{-\kappa_-}{2} \hat{a}_-(t)
		+ i g_- \left[ \hat{b}(t) e^{-i (\omega_m + \Delta_-) t} + \hat{b}^{\dagger}(t) e^{i (\omega_m-\Delta_-) t} \right]
		+ \sqrt{\kappa_\mathrm{ex, -}} \hat{a}_\mathrm{in}(t) e^{-i\Delta_- t}, \\
		\frac{d}{dt} \hat{a}_+(t) = &\frac{-\kappa_+}{2} \hat{a}_+(t)
		+ i g_+ \left[ \hat{b}(t) e^{-i (\omega_m + \Delta_+) t} + \hat{b}^{\dagger}(t) e^{i (\omega_m-\Delta_+) t} \right]
		+ \sqrt{\kappa_\mathrm{ex, +}} \hat{a}_\mathrm{in}(t) e^{-i\Delta_+ t}, \\
		\frac{d}{dt} \hat{b}(t) = &\frac{-\kappa_m}{2} \hat{b}(t)
		+ i g_-^* \hat{a}_-(t) e^{i (\omega_m + \Delta_-) t}
		+ i g_- \hat{a}_-^{\dagger}(t) e^{i (\omega_m - \Delta_-) t} \\
		&+ i g_+^* \hat{a}_+(t) e^{i (\omega_m + \Delta_+) t}
		+ i g_+ \hat{a}_+^{\dagger}(t) e^{i (\omega_m - \Delta_+) t}
		+ \sqrt{\kappa_\mathrm{ex, m}} \hat{c}_\mathrm{in}(t).
	\end{aligned}
\end{equation}
We can move to the frequency domain by using the Fourier transform
\begin{equation}
	\hat{a}[\omega] = \mathcal{F}\left( \hat{a}(t) \right)[\omega] = \int_{-\infty}^{\infty} \hat{a}(t) e^{i \omega t} dt
\end{equation}
along with the properties 
\begin{equation}
	\begin{aligned}
		\mathcal{F}\left( \frac{d}{dt} \hat{a}(t) \right)[\omega] &= - i \omega \hat{a}[\omega],\\
		\mathcal{F}\left( \hat{a}(t) e^{-i \Delta t} \right)[\omega] &= \hat{a}[\omega - \Delta],\\
		\hat{a}^{\dagger}[\omega] \equiv \mathcal{F}\left( \hat{a}^{\dagger}(t) \right)[\omega] &= \left( \hat{a}[-\omega] \right)^{\dagger}.
	\end{aligned}
\end{equation}
We then have
\begin{equation}\label{eq:langevin}
	\begin{aligned}
		\hat{a}_-[\omega] =\ &i g_- \chi_-[\omega] \left( \hat{b}[\omega - (\omega_m + \Delta_-)] + \hat{b}^{\dagger}[\omega + (\omega_m-\Delta_-) ] \right)
		+ \sqrt{\kappa_\mathrm{ex, -}} \chi_-[\omega] \hat{a}_\mathrm{in}[\omega - \Delta_-], \\
		\hat{a}_+[\omega] =\ &i g_+ \chi_+[\omega] \left( \hat{b}[\omega - (\omega_m + \Delta_+)]  + \hat{b}^{\dagger}[\omega + (\omega_m-\Delta_+)] \right)
		+ \sqrt{\kappa_\mathrm{ex, +}} \chi_+[\omega] \hat{a}_\mathrm{in}[\omega - \Delta_+], \\
		\hat{b}[\omega] =\ &i g_-^* \chi_m[\omega] \hat{a}_-[\omega + (\omega_m + \Delta_-)]
		+ i g_- \chi_m[\omega] \hat{a}_-^{\dagger}[\omega + (\omega_m - \Delta_-)] \\
		&+ i g_+^* \chi_m[\omega] \hat{a}_+[\omega + (\omega_m + \Delta_+)] 
		+\ i g_+ \chi_m[\omega] \hat{a}_+^{\dagger}[\omega + (\omega_m - \Delta_+)]
		+ \sqrt{\kappa_\mathrm{ex, m}} \chi_m[\omega] \hat{c}_\mathrm{in}[\omega],
	\end{aligned}
\end{equation}
with the susceptibilities $\chi_-[\omega] = \left( -i\omega + \kappa_-/2 \right)^{-1}$, $\chi_+[\omega] = \left( -i\omega + \kappa_+/2 \right)^{-1}$ and $\chi_m[\omega] = \left( -i\omega + \kappa_m/2 \right)^{-1}$.
The bath operators are linked through the input-output relations
\begin{equation}
	\begin{aligned}
		\hat{a}_\mathrm{out}(t)e^{-i\omega_L t} &= \hat{a}_\mathrm{in}(t)e^{-i\omega_L t} - \sqrt{\kappa_{\mathrm{ex}, -}} \hat{a}_-(t)e^{-i\omega_- t} - \sqrt{\kappa_{\mathrm{ex}, +}} \hat{a}_+(t)e^{-i\omega_+ t}, \\
		\hat{c}_\mathrm{out}(t)e^{-i\omega_m t} &= -\hat{c}_\mathrm{in}(t)e^{-i\omega_m t} + \sqrt{\kappa_{\mathrm{ex}, m}} \hat{b}(t)e^{-i\omega_m t} 
	\end{aligned}
\end{equation}
or
\begin{equation}\label{eq:input-output}
	\begin{aligned}
		\hat{a}_\mathrm{out}[\omega] &= \hat{a}_\mathrm{in}[\omega] - \sqrt{\kappa_{\mathrm{ex}, -}} \hat{a}_-[\omega + \Delta_-] - \sqrt{\kappa_{\mathrm{ex}, +}} \hat{a}_+[\omega+\Delta_+], \\
		\hat{c}_\mathrm{out}[\omega] &= -\hat{c}_\mathrm{in}[\omega] + \sqrt{\kappa_{\mathrm{ex}, m}} \hat{b}[\omega] 
	\end{aligned}
\end{equation}
in frequency domain.
They satisfy the commutation relations $\left[ \hat{a}_\mathrm{in}[\omega], \hat{a}_\mathrm{in}^{\dagger}[\omega'] \right] = \delta[\omega-\omega']$ and $\left[ \hat{c}_\mathrm{in}[\omega], \hat{c}_\mathrm{in}^{\dagger}[\omega'] \right] = \delta[\omega-\omega']$.

\subsection{Conversion efficiency and added noise}\label{supp:transfer_functions}
\subsubsection{Anti-Stokes configuration}
For the following analysis, we consider the triply resonant configuration where $\omega_+ - \omega_-  = \omega_m$ and the optical pump is on resonance with one of the optical supermodes.
First, we consider the anti-Stokes configuration where $\Delta_- = 0$ and $\Delta_+ = -\omega_m$.
From Eq.~\ref{eq:input-output}, we have the input-output relation
\begin{equation}
	\hat{a}_\mathrm{out}[\omega] = \hat{a}_\mathrm{in}[\omega] - \sqrt{\kappa_{\mathrm{ex}, -}} \hat{a}_-[\omega] - \sqrt{\kappa_{\mathrm{ex}, +}} \hat{a}_+[\omega-\omega_m].
\end{equation}
The equations of motion is given by
\begin{equation}
	\begin{aligned}
		\hat{a}_-[\omega] &=\ i g_- \chi_-[\omega] \left( \hat{b}[\omega - \omega_m ] + \hat{b}^{\dagger}[\omega + \omega_m ] \right)
		+ \sqrt{\kappa_\mathrm{ex, -}} \chi_-[\omega] \hat{a}_\mathrm{in}[\omega]\\
		&\approx \sqrt{\kappa_\mathrm{ex, -}} \chi_-[\omega] \hat{a}_\mathrm{in}[\omega], \\
		\hat{a}_+[\omega] &=\ i g_+ \chi_+[\omega] \left( \hat{b}[\omega]  + \hat{b}^{\dagger}[\omega + 2 \omega_m] \right)
		+ \sqrt{\kappa_\mathrm{ex, +}} \chi_+[\omega] \hat{a}_\mathrm{in}[\omega + \omega_m] \\
		&\approx\ i g_+ \chi_+[\omega] \hat{b}[\omega]
		+ \sqrt{\kappa_\mathrm{ex, +}} \chi_+[\omega] \hat{a}_\mathrm{in}[\omega + \omega_m],\\
		\hat{b}[\omega] &=\ i g_-^* \chi_m[\omega] \hat{a}_-[\omega + \omega_m]
		+ i g_- \chi_m[\omega] \hat{a}_-^{\dagger}[\omega + \omega_m] 
		+ i g_+^* \chi_m[\omega] \hat{a}_+[\omega]
		+ i g_+ \chi_m[\omega] \hat{a}_+^{\dagger}[\omega + 2 \omega_m]
		+ \sqrt{\kappa_\mathrm{ex, m}} \chi_m[\omega] \hat{c}_\mathrm{in}[\omega]\\
		&\approx \ i g_+^* \chi_m[\omega] \hat{a}_+[\omega]
		+ \sqrt{\kappa_\mathrm{ex, m}} \chi_m[\omega] \hat{c}_\mathrm{in}[\omega].
	\end{aligned}
\end{equation}
Here we drop the terms that contain susceptibilities and ladder operators with an offsetted frequency dependence (with the exception of the bath operators), since they necessarily imply an attenuation of at least $\kappa_+ / \left( 2 \omega_m \right)$ compared to the dominant terms. 
This approximation is valid in the resolved-sideband regime $\kappa_-, \kappa_+ \ll \omega_m$.
Supplementary~Figure~\ref{suppfig:sfg}a shows the signal flow graph representing the equations of motion, from which we deduced the transfer functions between the microwave field at $\omega_m$ and the optical field at $\omega_L + \omega_m$ using Mason's rule \cite{blesinQuantumCoherentMicrowaveoptical2021}
\begin{equation}
	\begin{aligned}
		S_{\hat{a}_\mathrm{out}[\omega + \omega_m] \leftarrow \hat{c}_\mathrm{in}[\omega]}
		= &S_{\hat{c}_\mathrm{out}[\omega] \leftarrow \hat{a}_\mathrm{in}[\omega + \omega_m]}
		= S_{\hat{a}_\mathrm{out}^\dagger[\omega + \omega_m] \leftarrow \hat{c}_\mathrm{in}^\dagger[\omega]}^*
		= S_{\hat{c}_\mathrm{out}^\dagger[\omega] \leftarrow \hat{a}_\mathrm{in}^\dagger[\omega + \omega_m]}^*\\
		= &\frac{ -i \sqrt{ \kappa_\mathrm{ex, +} } \sqrt{ \kappa_\mathrm{ex, m} } g_+ \chi_+[\omega] \chi_m[\omega] }{1 + |g_+|^2 \chi_+[\omega] \chi_m[\omega]}.
	\end{aligned}
\end{equation}
The optical-optical and microwave-microwave transfer functions can also be derived as
\begin{align}
	S_{\hat{a}_\mathrm{out}[\omega + \omega_m] \leftarrow \hat{a}_\mathrm{in}[\omega + \omega_m]}
	&= 1 
	- \frac{ \kappa_\mathrm{ex, +} \chi_+[\omega] }{1 + |g_+|^2 \chi_+[\omega] \chi_m[\omega]}
	- \kappa_\mathrm{ex, -} \chi_-[\omega + \omega_m],\\
	S_{\hat{c}_\mathrm{out}[\omega] \leftarrow \hat{c}_\mathrm{in}[\omega]}
	&= -1 
	+ \frac{ \kappa_\mathrm{ex, m} \chi_m[\omega] }{1 + |g_+|^2 \chi_+[\omega] \chi_m[\omega]}\label{eq:TransferFunctionMicrowave}.
\end{align}
The other transfer functions (e.g., $S_{\hat{a}_\mathrm{out}^{\dagger}[\omega + \omega_m] \leftarrow \hat{c}_\mathrm{in}[\omega]}$) are expected to be small given the minuscule weight of the corresponding edges in the signal flow graph.
The photon number flux at the optical output of the transducer is given by
\begin{equation}
	\begin{aligned}
		\hat{a}_{\mathrm{out}}^{\dagger}[\omega + \omega_m]
		\hat{a}_{\mathrm{out}}[\omega + \omega_m]
		\approx&
		\left( S_{\hat{a}_{\mathrm{out}}[\omega + \omega_m] \leftarrow \hat{a}_{\mathrm{in}}[\omega + \omega_m] }^* \hat{a}_{\mathrm{in}}^{\dagger}[\omega + \omega_m]
		+ S_{\hat{a}_{\mathrm{out}}[\omega + \omega_m] \leftarrow \hat{c}_{\mathrm{in}}[\omega] }^* \hat{c}_{\mathrm{in}}^{\dagger}[\omega] \right) \\
		\ &\left( S_{\hat{a}_{\mathrm{out}}[\omega + \omega_m] \leftarrow \hat{a}_{\mathrm{in}}[\omega + \omega_m] } \hat{a}_{\mathrm{in}}[\omega + \omega_m]
		+ S_{\hat{a}_{\mathrm{out}}[\omega + \omega_m] \leftarrow \hat{c}_{\mathrm{in}}[\omega] } \hat{c}_{\mathrm{in}}[\omega] \right).
	\end{aligned}
\end{equation}
To capture the noise properties of the optical output, one has to measure its symmetrized power spectral density
\begin{equation}
	\begin{aligned}
		\mathcal{S}_{\hat{a}_{\mathrm{out}}}[\omega + \omega_m]
		=&\ \frac{1}{2} \expval{ \left\{ \hat{a}_{\mathrm{out}}[\omega + \omega_m], \hat{a}_{\mathrm{out}}^{\dagger}[\omega + \omega_m] \right\} }
		= \frac{1}{2} + \expval{\hat{a}_{\mathrm{out}}^{\dagger}[\omega + \omega_m]\hat{a}_{\mathrm{out}}[\omega + \omega_m]} \\
		=&\ \eta_+[\omega] \left( \mathcal{S}_{\hat{c}_{\mathrm{in}}}[\omega] + \mathcal{S}_\mathrm{added, +}^\uparrow[\omega] \right),
	\end{aligned}
\end{equation}
where $\mathcal{S}_{\hat{c}_{\mathrm{in}}}[\omega] = \frac{1}{2} + \expval{\hat{c}_{\mathrm{in}}^{\dagger}[\omega]\hat{c}_{\mathrm{in}}[\omega]}$ is the power spectral density of the microwave photon flux at the input port.
%This allows defining the on-chip photon number conversion efficiency
We define the on-chip photon number transduction efficiency as
\begin{equation}
	\label{eq:OnChipEfficiencyAntiStokes}
	\begin{aligned}
		\eta_+[\omega]
		&= |S_{\hat{a}_\mathrm{out}[\omega + \omega_m] \leftarrow \hat{c}_\mathrm{in}[\omega]} |^2
		= |S_{\hat{c}_\mathrm{out}[\omega] \leftarrow \hat{a}_\mathrm{in}[\omega + \omega_m]} |^2 \\
		&= \frac{ \kappa_\mathrm{ex, +} \kappa_\mathrm{ex, m} |g_+|^2 |\chi_+[\omega]|^2 |\chi_m[\omega]|^2 }{|1 + |g_+|^2 \chi_+[\omega] \chi_m[\omega]|^2}
		\approx \frac{\kappa_\mathrm{ex, +}}{\kappa_+} \frac{\kappa_\mathrm{ex, m}}{\kappa_m} \frac{4 C_+}{(1+C_+)^2}
	\end{aligned}
\end{equation}
and the added noise during up-conversion as
\begin{equation}
	\begin{aligned}
		\mathcal{S}_\mathrm{added, +}^\uparrow[\omega] =&\ \frac{1}{2 \eta_+[\omega]} \left( 1 - \eta_+[\omega] - |S_{\hat{a}_{\mathrm{out}}[\omega + \omega_m] \leftarrow \hat{a}_{\mathrm{in}}[\omega + \omega_m] }|^2 \right)
		+ \frac{ |S_{\hat{a}_{\mathrm{out}}[\omega + \omega_m] \leftarrow \hat{a}_{\mathrm{in}}[\omega + \omega_m] }|^2 }{ \eta_+[\omega] } \  \mathcal{S}_{\hat{a}_{\mathrm{in}}}[\omega + \omega_m].
	\end{aligned}
\end{equation}
We used the fact that the noise from the two input ports are not correlated, i.e. $ \expval{ \hat{c}_{\mathrm{in}}^{\dagger}[\omega] \hat{a}_{\mathrm{in}}[\omega + \omega_m] } 
= \expval{ \hat{c}_{\mathrm{in}}[\omega] \hat{a}_{\mathrm{in}}^{\dagger}[\omega + \omega_m] } 
= 0 $. 
An additional noise of $\frac{1}{2 \eta_+[\omega]} \left( 1 - \eta_+[\omega] - |S_{\hat{a}_{\mathrm{out}}[\omega + \omega_m] \leftarrow \hat{a}_{\mathrm{in}}[\omega + \omega_m] }|^2 \right)$ is introduced so that the output channel still respect the bosonic commutation relations \cite{caves_quantum_1982}.
%\begin{equation}
%	\begin{aligned}
%		\hat{n}_\mathrm{added, +}^\uparrow[\omega] =&\ 
%		\frac{ S_{\hat{a}_{\mathrm{out}}[\omega + \omega_m] \leftarrow \hat{a}_{\mathrm{in}}[\omega + \omega_m] }^* }{ S_{\hat{a}_{\mathrm{out}}[\omega + \omega_m] \leftarrow \hat{c}_{\mathrm{in}}[\omega] }^* }  \hat{a}_{\mathrm{in}}^{\dagger}[\omega + \omega_m] \hat{c}_{\mathrm{in}}[\omega] 
%		+ \frac{ S_{\hat{a}_{\mathrm{out}}[\omega + \omega_m] \leftarrow \hat{a}_{\mathrm{in}}[\omega + \omega_m] } }{ S_{\hat{a}_{\mathrm{out}}[\omega + \omega_m] \leftarrow \hat{c}_{\mathrm{in}}[\omega] } }  \hat{c}_{\mathrm{in}}^{\dagger}[\omega] \hat{a}_{\mathrm{in}}[\omega + \omega_m] \\
%		&+ \frac{ |S_{\hat{a}_{\mathrm{out}}[\omega + \omega_m] \leftarrow \hat{a}_{\mathrm{in}}[\omega + \omega_m] }|^2 }{ |S_{\hat{a}_{\mathrm{out}}[\omega + \omega_m] \leftarrow \hat{c}_{\mathrm{in}}[\omega] }|^2 } \  \hat{a}_{\mathrm{in}}^{\dagger}[\omega + \omega_m] \hat{a}_{\mathrm{in}}[\omega + \omega_m].
%	\end{aligned}
%\end{equation}
The last approximation of Eq.~\ref{eq:OnChipEfficiencyAntiStokes} gives the steady-state efficiency of the transducer ($\omega\rightarrow 0$), which depends on anti-Stokes cooperativity $C_+ = 4 g_+^2/\left(\kappa_+ \kappa_m\right)$ as well as extraction efficiencies of the antisymmetric supermode $\kappa_\mathrm{ex, +}/\kappa_+$ and the acoustic mode $\kappa_\mathrm{ex, m}/\kappa_m$.
Similarly, the symmetrized power spectral density of the output photon number flux at the microwave port is
\begin{equation}
	\begin{aligned}
		\mathcal{S}_{\hat{c}_{\mathrm{out}}}[\omega]
		=&\ \frac{1}{2} \expval{ \left\{ \hat{c}_{\mathrm{out}}[\omega ], \hat{c}_{\mathrm{out}}^{\dagger}[\omega] \right\} }
		= \frac{1}{2} + \expval{\hat{c}_{\mathrm{out}}^{\dagger}[\omega]\hat{c}_{\mathrm{out}}[\omega]} 
		=\ \eta_+[\omega] \left( \mathcal{S}_{\hat{a}_{\mathrm{in}}}[\omega+\omega_m] + \mathcal{S}_\mathrm{added, +}^\downarrow[\omega] \right).
	\end{aligned}
\end{equation}
%%	\begin{aligned}
%		\hat{c}_{\mathrm{out}}^{\dagger}
%		\hat{c}_{\mathrm{out}}[\omega]
%		\approx&\ \eta_+[\omega] \left( \hat{a}_{\mathrm{in}}^{\dagger} \hat{a}_{\mathrm{in}}[\omega + \omega_m] + \hat{n}_\mathrm{added, +}^\downarrow[\omega] \right).
%	\end{aligned}
%\end{equation}
While the expressions for the conversion efficiency are identical for both up- and down-conversion, the added noise for down-conversion is instead
\begin{equation}
	\begin{aligned}
		\mathcal{S}_\mathrm{added, +}^\downarrow[\omega] =&\ \frac{1}{2 \eta_+[\omega]} \left( 1 - \eta_+[\omega] - |S_{\hat{c}_{\mathrm{out}}[\omega] \leftarrow \hat{c}_{\mathrm{in}}[\omega] }|^2 \right)
		+\frac{ |S_{\hat{c}_{\mathrm{out}}[\omega] \leftarrow \hat{c}_{\mathrm{in}}[\omega] }|^2 }{ \eta_+[\omega] } \  \mathcal{S}_{\hat{c}_{\mathrm{in}}}[\omega].
	\end{aligned}
\end{equation}
%\begin{equation}
%	\begin{aligned}
%		\hat{n}_\mathrm{added, +}^\downarrow[\omega] =&\ 
%		\frac{ S_{\hat{c}_{\mathrm{out}}[\omega] \leftarrow \hat{c}_{\mathrm{in}}[\omega] } }{ S_{\hat{c}_{\mathrm{out}}[\omega] \leftarrow \hat{a}_{\mathrm{in}}[\omega + \omega_m] } }  \hat{a}_{\mathrm{in}}^{\dagger}[\omega + \omega_m] \hat{c}_{\mathrm{in}}[\omega] 
%		+ \frac{ S_{\hat{c}_{\mathrm{out}}[\omega] \leftarrow \hat{c}_{\mathrm{in}}[\omega] }^* }{ S_{\hat{c}_{\mathrm{out}}[\omega] \leftarrow \hat{a}_{\mathrm{in}}[\omega + \omega_m] }^* } \hat{c}_{\mathrm{in}}^{\dagger}[\omega] \hat{a}_{\mathrm{in}}[\omega + \omega_m] \\
%		&+\frac{ |S_{\hat{c}_{\mathrm{out}}[\omega] \leftarrow \hat{c}_{\mathrm{in}}[\omega] }|^2 }{ |S_{\hat{c}_{\mathrm{out}}[\omega] \leftarrow \hat{a}_{\mathrm{in}}[\omega + \omega_m] }|^2 } \  \hat{c}_{\mathrm{in}}^{\dagger}[\omega] \hat{c}_{\mathrm{in}}[\omega].
%	\end{aligned}
%\end{equation}

\subsubsection{Stokes configuration}
We follow the same recipe to derive the conversion efficiency and added noise for the Stokes configuration where $\Delta_- = \omega_m$ and $\Delta_+ = 0$. 
We have the optical input-output relation
\begin{equation}
	\hat{a}_\mathrm{out}[\omega] = \hat{a}_\mathrm{in}[\omega] - \sqrt{\kappa_{\mathrm{ex}, -}} \hat{a}_-[\omega + \omega_m] - \sqrt{\kappa_{\mathrm{ex}, +}} \hat{a}_+[\omega]
\end{equation}
and the equations of motion
\begin{equation}
	\begin{aligned}
		\hat{a}_-[\omega] &=\ i g_- \chi_-[\omega] \left( \hat{b}[\omega - 2\omega_m] + \hat{b}^{\dagger}[\omega] \right)
		+ \sqrt{\kappa_\mathrm{ex, -}} \chi_-[\omega] \hat{a}_\mathrm{in}[\omega - \omega_m] \\
		&\approx \ i g_- \chi_-[\omega] \hat{b}^{\dagger}[\omega]
		+ \sqrt{\kappa_\mathrm{ex, -}} \chi_-[\omega] \hat{a}_\mathrm{in}[\omega - \omega_m] , \\
		\hat{a}_+[\omega] &=\ i g_+ \chi_+[\omega] \left( \hat{b}[\omega - \omega_m]  + \hat{b}^{\dagger}[\omega + \omega_m] \right)
		+ \sqrt{\kappa_\mathrm{ex, +}} \chi_+[\omega] \hat{a}_\mathrm{in}[\omega] \\
		&\approx \ \sqrt{\kappa_\mathrm{ex, +}} \chi_+[\omega] \hat{a}_\mathrm{in}[\omega] , \\
		\hat{b}[\omega] &=\ i g_-^* \chi_m[\omega] \hat{a}_-[\omega + 2\omega_m]
		+ i g_- \chi_m[\omega] \hat{a}_-^{\dagger}[\omega]
		+ i g_+^* \chi_m[\omega] \hat{a}_+[\omega + \omega_m]
		+ i g_+ \chi_m[\omega] \hat{a}_+^{\dagger}[\omega + \omega_m]
		+ \sqrt{\kappa_\mathrm{ex, m}} \chi_m[\omega] \hat{c}_\mathrm{in}[\omega] \\
		&\approx \ i g_- \chi_m[\omega] \hat{a}_-^{\dagger}[\omega] + \sqrt{\kappa_\mathrm{ex, m}} \chi_m[\omega] \hat{c}_\mathrm{in}[\omega].
	\end{aligned}
\end{equation}
Using the signal flow graph shown in Supplementary~Fig.~\ref{suppfig:sfg}b, the transfer functions between the microwave field at $\omega_m$ and the optical field at $\omega_L - \omega_m$ can once again be derived as
\begin{align}
	S_{\hat{a}_\mathrm{out}[\omega - \omega_m] \leftarrow \hat{c}_\mathrm{in}^{\dagger}[\omega]}
	&= S_{\hat{a}_\mathrm{out}^{\dagger}[\omega-\omega_m] \leftarrow \hat{c}_\mathrm{in}[\omega]}^*
	= \frac{ -i \sqrt{ \kappa_\mathrm{ex, -} } \sqrt{ \kappa_\mathrm{ex, m} } g_- \chi_-[\omega] \chi_m^*[\omega] }{1 - |g_-|^2 \chi_-^*[\omega] \chi_m[\omega]},\\
	S_{\hat{c}_\mathrm{out}[\omega] \leftarrow \hat{a}_\mathrm{in}^{\dagger}[\omega - \omega_m]}
	&= S_{\hat{c}_\mathrm{out}^{\dagger}[\omega] \leftarrow \hat{a}_\mathrm{in}[\omega - \omega_m]}^*
	= \frac{ i \sqrt{ \kappa_\mathrm{ex, -} } \sqrt{ \kappa_\mathrm{ex, m} } g_- \chi_-^*[\omega] \chi_m[\omega] }{1 - |g_-|^2 \chi_-[\omega] \chi_m^*[\omega]}.
\end{align}
While the microwave-microwave transfer function is identical to Eq.~\ref{eq:TransferFunctionMicrowave}, the optical-optical transfer function is
\begin{equation}
	S_{\hat{a}_\mathrm{out}[\omega - \omega_m] \leftarrow \hat{a}_\mathrm{in}[\omega - \omega_m]}
	= 1 
	- \frac{ \kappa_\mathrm{ex, -} \chi_-[\omega] }{1 - |g_-|^2 \chi_-^*[\omega] \chi_m[\omega]}
	- \kappa_\mathrm{ex, +} \chi_+[\omega - \omega_m].
\end{equation}
%The optical photon number at the output of the transducer at $\omega_L - \omega_m$ can be computed as
%\begin{equation}
%	\label{eq:StokesOpticalOutput}
%	\begin{aligned}
%		\hat{a}_{\mathrm{out}}^{\dagger}
%		\hat{a}_{\mathrm{out}}[\omega - \omega_m]
%		\approx&\ \eta_-[\omega] \left( \hat{c}_{\mathrm{in}}^{\dagger} \hat{c}_{\mathrm{in}}[\omega] + \hat{n}_\mathrm{added, -}^\uparrow[\omega] \right),
%	\end{aligned}
%\end{equation}
The photon number flux at the optical output of the transducer has a symmetrized power spectral density
\begin{equation}
	\label{eq:StokesOpticalOutput}
	\begin{aligned}
		\mathcal{S}_{\hat{a}_{\mathrm{out}}}[\omega - \omega_m]
		=&\ \frac{1}{2} \expval{ \left\{ \hat{a}_{\mathrm{out}}[\omega - \omega_m], \hat{a}_{\mathrm{out}}^{\dagger}[\omega - \omega_m] \right\} }
		= \frac{1}{2} + \expval{\hat{a}_{\mathrm{out}}^{\dagger}[\omega - \omega_m]\hat{a}_{\mathrm{out}}[\omega - \omega_m]} \\
		=&\ \eta_-[\omega] \left( \mathcal{S}_{\hat{c}_{\mathrm{in}}}[\omega] + \mathcal{S}_\mathrm{added, -}^\uparrow[\omega] \right).
	\end{aligned}
\end{equation}
We define the on-chip photon number conversion efficiency for the Stokes configuration as
\begin{equation}
	\label{eq:OnChipEfficiencyStokes}
	\begin{aligned}
		\eta_-[\omega]
		&= |S_{\hat{a}_{\mathrm{out}}[\omega - \omega_m] \leftarrow \hat{c}_{\mathrm{in}}^{\dagger}[\omega] }|^2
		= |S_{\hat{c}_\mathrm{out}[\omega] \leftarrow \hat{a}_\mathrm{in}^{\dagger}[\omega - \omega_m]} |^2 \\
		&= \frac{ \kappa_\mathrm{ex, -} \kappa_\mathrm{ex, m} |g_-|^2 |\chi_-[\omega]|^2 |\chi_m[\omega]|^2 }{|1 + |g_-|^2 \chi_-[\omega] \chi_m[\omega]|^2}
		\approx \frac{\kappa_\mathrm{ex, -}}{\kappa_-} \frac{\kappa_\mathrm{ex, m}}{\kappa_m} \frac{4 C_-}{(1-C_-)^2}
	\end{aligned}
\end{equation}
as well as the added noise during up-conversion
\begin{equation}
	\begin{aligned}
		\mathcal{S}_\mathrm{added, -}^\uparrow[\omega] =&\ \frac{1}{2 \eta_-[\omega]} \left( 1 + \eta_-[\omega] - |S_{\hat{a}_{\mathrm{out}}[\omega - \omega_m] \leftarrow \hat{a}_{\mathrm{in}}[\omega - \omega_m] }|^2 \right)
		+ \frac{ |S_{\hat{a}_{\mathrm{out}}[\omega - \omega_m] \leftarrow \hat{a}_{\mathrm{in}}[\omega - \omega_m] }|^2 }{ \eta_-[\omega] } \  \mathcal{S}_{\hat{a}_{\mathrm{in}}}[\omega - \omega_m]
	\end{aligned}
\end{equation}
where $C_- = 4 g_-^2/\left(\kappa_- \kappa_m\right)$ is the Stokes cooperativity.
%Importantly, for a temperature $T$, Boltzmann's constant $k_B$ and number of thermal excitations in the microwave bath $n_\mathrm{m, th}(\omega) = \left[ \exp\left( \hbar \omega/(k_B T) \right)  - 1 \right]^{-1}$, the commutation relation $\left[ \hat{c}_{\mathrm{in}}[\omega], \hat{c}_{\mathrm{in}}^{\dagger}[\omega] \right] = 1 + n_\mathrm{m, th}(\omega)$ gives a fundamental limit on the added noise during up-conversion.
The output photon number at the microwave port for the Stokes configuration is given by
\begin{equation}
	\label{eq:StokesMicrowaveOutput}
	\begin{aligned}
		\mathcal{S}_{\hat{c}_{\mathrm{out}}}[\omega]
		=&\ \frac{1}{2} \expval{ \left\{ \hat{c}_{\mathrm{out}}[\omega ], \hat{c}_{\mathrm{out}}^{\dagger}[\omega] \right\} }
		= \frac{1}{2} + \expval{\hat{c}_{\mathrm{out}}^{\dagger}[\omega]\hat{c}_{\mathrm{out}}[\omega]} 
		=\ \eta_-[\omega] \left( \mathcal{S}_{\hat{a}_{\mathrm{in}}}[\omega-\omega_m] + \mathcal{S}_\mathrm{added, -}^\downarrow[\omega] \right).
	\end{aligned}
\end{equation}
%\begin{equation}
%	\label{eq:StokesMicrowaveOutput}
%	\begin{aligned}
%		\hat{c}_{\mathrm{out}}^{\dagger}
%		\hat{c}_{\mathrm{out}}[\omega]
%		\approx&\ \eta_-[\omega] \left( \hat{a}_{\mathrm{in}}^{\dagger} \hat{a}_{\mathrm{in}}[\omega - \omega_m] + \hat{n}_\mathrm{added, -}^\downarrow[\omega] \right).
%	\end{aligned}
%\end{equation}
The photon number on-chip conversion efficiency is again the same for both up- and down-conversion, but the added noise for the down-conversion process is instead
\begin{equation}
	\begin{aligned}
		\mathcal{S}_\mathrm{added, -}^\downarrow[\omega] =&\ \frac{1}{2 \eta_-[\omega]} \left( 1 + \eta_-[\omega] - |S_{\hat{c}_{\mathrm{out}}[\omega] \leftarrow \hat{c}_{\mathrm{in}}[\omega] }|^2 \right)
		+\frac{ |S_{\hat{c}_{\mathrm{out}}[\omega] \leftarrow \hat{c}_{\mathrm{in}}[\omega] }|^2 }{ \eta_-[\omega] } \  \mathcal{S}_{\hat{c}_{\mathrm{in}}}[\omega].
	\end{aligned}
\end{equation}
%\begin{equation}
%	\begin{aligned}
%		\hat{n}_\mathrm{added, -}^\downarrow[\omega] =&\ 
%		\frac{ S_{\hat{c}_{\mathrm{out}}[\omega] \leftarrow \hat{c}_{\mathrm{in}}[\omega] } }{ S_{\hat{c}_{\mathrm{out}}[\omega] \leftarrow \hat{a}_{\mathrm{in}}^{\dagger}[\omega - \omega_m] } }   \hat{a}_{\mathrm{in}}[\omega - \omega_m] \hat{c}_{\mathrm{in}}[\omega]
%		+ \frac{ S_{\hat{c}_{\mathrm{out}}[\omega] \leftarrow \hat{c}_{\mathrm{in}}[\omega] }^* }{ S_{\hat{c}_{\mathrm{out}}[\omega] \leftarrow \hat{a}_{\mathrm{in}}^{\dagger}[\omega - \omega_m] }^* } \hat{c}_{\mathrm{in}}^{\dagger}[\omega] \hat{a}_{\mathrm{in}}^{\dagger}[\omega - \omega_m] \\
%		&+ \frac{ |S_{\hat{c}_{\mathrm{out}}[\omega] \leftarrow \hat{c}_{\mathrm{in}}[\omega] }|^2 }{ |S_{\hat{c}_{\mathrm{out}}[\omega] \leftarrow \hat{a}_{\mathrm{in}}^{\dagger}[\omega - \omega_m] }|^2 } \  \hat{c}_{\mathrm{in}}^{\dagger}[\omega] \hat{c}_{\mathrm{in}}[\omega] 
%		+ 1.
%	\end{aligned}
%\end{equation}
%As in the up-conversion case, the commutation relation $\left[ \hat{a}_{\mathrm{in}}[\omega], \hat{a}_{\mathrm{in}}^{\dagger}[\omega] \right] = 1$ gives a fundamental limit on the added noise during down-conversion. 
%$\hat{n}_\mathrm{added, -}^\downarrow[\omega] \geq 1$.

\subsection{Estimation of correlated microwave-optical pair generation rates}\label{supp:pair_rates}
Equations~\ref{eq:StokesOpticalOutput} and \ref{eq:StokesMicrowaveOutput} suggest that entangled pairs of microwave and optical photons can be generated in the Stokes configuration since it gives rise to an effective two-mode squeezing Hamiltonian \cite{ruedaElectroopticEntanglementSource2019,zhongProposalHeraldedGeneration2020,krastanovOpticallyHeraldedEntanglement2021}.
In order to verify that the HBAR transducer can generate entangled pairs of microwave and optical photons, we compute the on-chip pair generation rate, expected to be equal to the output photon flux from the lower frequency optical mode as in standard spontaneous parametric down-conversion (SPDC) scheme.
Integrating over the transduction bandwidth, we have
\begin{equation}
	R = \int_{-\infty}^{\infty} \expval{\hat{a}_\mathrm{out}^{\dagger}[\omega] \hat{a}_\mathrm{out}[\omega]} d\omega
	\approx \int_{-\infty}^{\infty} \eta_-[\omega] d\omega
	\approx \frac{\pi}{2} \left( \kappa_-^{-1} + \kappa_m^{-1} \right)^{-1} \eta_-[0].
\end{equation}
In practice, the pump laser should be filtered for the optical output to be dominated by heralding photons, and other losses from the measurement setup can further reduce the effective rate.
For $\eta^\mathrm{oc}=\SI{7.9e-5}{}$, $\kappa_- = 2\pi\times\SI{166}{\mega\hertz}$ and $\kappa_m = 2\pi\times\SI{13}{\mega\hertz}$, this would corresponds to an on-chip pair generation rate of $R \approx 2\pi \times \SI{1.5}{\kilo\hertz}$.
Losses in the measurement chain have to be accounted for to estimate the count rate at the detector.
In order to measure entanglement between the microwave and optical fields, the on-chip rate must be higher than the thermal decoherence rate of the acoustic mode $\Gamma^\mathrm{dec} = \kappa_m n_\mathrm{th, m}(\omega_m)$, where the thermal occupancy is given by the Bose-Einstein distribution $n_\mathrm{th, m}(\omega) = \left[ \exp\left( \hbar \omega/(k_B T) \right)  - 1 \right]^{-1}$ given a temperature $T$ and Boltzmann's constant $k_B$.
%In order to measure entanglement between the microwave and optical fields, the on-chip rate must be higher than the thermal decoherence rate of the acoustic mode $\Gamma^\mathrm{dec} = \kappa_m n_\mathrm{th, m}(\omega_m)$, where the thermal occupancy $n_\mathrm{th, m}(\omega_m)$ is given by the Bose-Einstein distribution.
For $\omega_m=2\pi\times\SI{3.5}{\giga\hertz}$ and $\kappa_m=2\pi\times\SI{13}{\mega\hertz}$, the decoherence rate $\Gamma^\mathrm{dec}/(2\pi)$ at \SI{800}{\milli\kelvin} is \SI{56}{\mega\hertz} and \SI{0.7}{\hertz} at \SI{10}{\milli\kelvin}.
Therefore, the transducer needs to be operated in the mixing chamber of the dilution refrigerator, with a pulsed pump to reduce heat load.
The total rate of optical heralding is hence gated by the pump duty cycle.

To evaluate the non-classical correlation required for the DLCZ protocol \cite{duanLongdistanceQuantumCommunication2001,krastanovOpticallyHeraldedEntanglement2021}, one can compute the second-order cross-correlation function.
Since no input signal needs to be sent to the transducer for SPDC, 
$\expval{ \hat{a}_\mathrm{in} } = \expval{ \hat{a}_\mathrm{in}^\dagger } = \expval{ \hat{a}_\mathrm{in}^\dagger \hat{a}_\mathrm{in} } = \expval{ \hat{a}_\mathrm{in} \hat{a}_\mathrm{in} } = \expval{ \hat{a}_\mathrm{in}^\dagger \hat{a}_\mathrm{in}^\dagger } = 0$ and $\expval{ \hat{c}_\mathrm{in}^\dagger[\omega] \hat{c}_\mathrm{in}[\omega] } = n_\mathrm{th, m}(\omega)$, leading to
\begin{equation}
	\begin{aligned}
		g_{ac}^{(2)} 
		= &\frac{ \expval{ \hat{c}_\mathrm{out}^\dagger \hat{a}_\mathrm{out}^\dagger \hat{a}_\mathrm{out} \hat{c}_\mathrm{out} } }{ \expval{\hat{c}_\mathrm{out}^\dagger \hat{c}_\mathrm{out}} \expval{\hat{a}_\mathrm{out}^\dagger \hat{a}_\mathrm{out}} } \\
		\approx &\frac{ \left( |S_{\hat{a}_{\mathrm{out}} \leftarrow \hat{a}_{\mathrm{in}} }|^2 + \eta_- \right) }
		{ \left( \eta_- + |S_{\hat{c}_{\mathrm{out}} \leftarrow \hat{c}_{\mathrm{in}} }|^2 n_\mathrm{th, m} \right) \left( 1 +  n_\mathrm{th, m} \right) } + \frac{ n_\mathrm{th, m} }{  1 +  n_\mathrm{th, m} } \\
		&+ \frac{ \left( S_{\hat{a}_{\mathrm{out}} \leftarrow \hat{a}_{\mathrm{in}} }^* S_{\hat{a}_{\mathrm{out}} \leftarrow \hat{c}_{\mathrm{in}}^{\dagger} } S_{\hat{c}_{\mathrm{out}} \leftarrow \hat{a}_{\mathrm{in}}^{\dagger} }^* S_{\hat{c}_{\mathrm{out}} \leftarrow \hat{c}_{\mathrm{in}} }
		+ S_{\hat{a}_{\mathrm{out}} \leftarrow \hat{a}_{\mathrm{in}} } S_{\hat{a}_{\mathrm{out}} \leftarrow \hat{c}_{\mathrm{in}}^{\dagger} }^* S_{\hat{c}_{\mathrm{out}} \leftarrow \hat{a}_{\mathrm{in}}^{\dagger} } S_{\hat{c}_{\mathrm{out}} \leftarrow \hat{c}_{\mathrm{in}} }^* \right) n_\mathrm{th, m} }
		{ \eta_- \left( \eta_- + |S_{\hat{c}_{\mathrm{out}} \leftarrow \hat{c}_{\mathrm{in}} }|^2 n_\mathrm{th, m} \right) \left( 1 +  n_\mathrm{th, m} \right) }
	\end{aligned}
\end{equation}
where the frequency dependences are omitted for readability.
Non-classicality is expected to lead to violation of the Cauchy-Schwarz inequality, given by $g_{ac}^{(2)} \le \sqrt{g_{aa}^{(2)}g_{cc}^{(2)}}$.
Therefore, it is desirable to minimize both microwave thermal occupancy (cooling) and optical transmission (attaining critical coupling).

\section{Fabrication process flow}\label{supp:process_flow}
The piezoelectric actuators are monolithically integrated on \ch{Si3N4} waveguides, fabricated using the photonic Damascene process \cite{pfeifferUltrasmoothSiliconNitride2018,liuHighyieldWaferscaleFabrication2021}.
The \SIadj{850}{\nano\meter}-thick \ch{Si3N4} film is deposited using low-pressure chemical vapor deposition into the Damascene preform, with \SI{2.8}{\micro\meter} thermal oxide below the waveguides.
After annealing to drive hydrogen impurities out of the \ch{Si3N4} layer, \SIadj{2.1}{\micro\meter}-thick TEOS oxide and \SIadj{1.0}{\micro\meter}-thick low-temperature oxide top claddings are deposited and subsequently annealed.
The metallic and piezoelectric films---\SI{95}{\nano\meter} of \ch{Mo}, \SI{1.0}{\micro\meter} of \ch{AlN} and another \SI{95}{\nano\meter} of \ch{Mo}---are sputtered through foundry services provided by Plasma-Therm \cite{tianHybridIntegratedPhotonics2020}.
They are then patterned using deep reactive ion etching (DRIE) to form the actuators, ground planes and integrated heaters.
We employ the same Band-Aid process as in Ref.~\cite{siddharthHertzlinewidthFrequencyagilePhotonic2023} to localize the electrodes atop the suspended cladding.
In particular, when connecting the \ch{Al} feedline to the top electrode by a lift-off process, the bottom electrode is etched back using \ch{XeF2} to avoid short-circuiting the electrodes.
The process to suspend the cladding in the vicinity of the actuator is combined with the chip singulation steps of the Damascene process. 
First, a hole is opened in the middle of the donut-shaped actuator. 
The chip facets are simultaneously defined where trenches in the \ch{SiO2} are etched by \ch{C4F8} DRIE between neighboring chips. 
A second photolithography is performed to protect the facets while leaving the etched holes exposed for further processing, resulting in narrower trenches between chips.
The \ch{Si} substrate is then isotropically etched with \ch{SF6} until the \ch{SiO2} cladding below the actuator is suspended \cite{tianMagneticfreeSiliconNitride2021}. 
Following cladding suspension, the \ch{Si} substrate is etched anisotropically until the desired chip thickness of \SI{250}{\micro\meter} is reached. 
A final isotropic \ch{Si} etch removes the parts of the substrate protruding from the \ch{SiO2} facet,
providing proper access to the bus waveguide nanotapers with lensed fibers, or even using the butt-coupling scheme. 
To facilitate this last step, we employ ion-beam etching at a $20^\circ$ tilt angle to remove the passivation layer formed on the hole sidewalls during the \ch{Si} DRIE.
The chips are ultimately released by grinding the back side of the wafer. 
This convoluted process ensures that all the actuators on the wafer are fully suspended, while the chip facets do not become exceedingly fragile from the undercut.

\section{Optical characterization}\label{supp:optical_characterization}

\subsection{Resonance linewidths}\label{supp:linewidth}
Fitting the hybridized transmission spectrum in Fig.~\ref{fig:fig_2}c to the coupled mode theory presented in Appendix~\ref{supp:hybrid} and in Ref.~\cite{blesinQuantumCoherentMicrowaveoptical2021} yields $\kappa_{r}/(2\pi) = \SI{154}{\mega\hertz}$ and $\kappa_{l}/(2\pi) = \SI{190}{\mega\hertz}$, with $\kappa_{\mathrm{ex}, l}/(2\pi) = \kappa_{\mathrm{ex}, r}/(2\pi) = \SI{60}{\mega\hertz}$.

\subsection{Hybridization of optical resonances via integrated thermo-optic heaters}\label{supp:heaters}
For the purpose of rapid characterization of the photonic molecules at room temperature,
the bottom electrode of the piezoelectric actuator also serves as an integrated heater to control the relative detuning between the micro-rings by thermo-optic effect.
As seen in Fig.~\ref{fig:fig_2}a, we pattern \ch{Mo} to make three electrical connections to the bottom electrode. 
The top and bottom connections are connected to ground, whereas the central connection is biased to a constant voltage.
The finite resistance of \ch{Mo} at room temperature leads to Joule heating as current flows through the bottom electrode, modifying the refractive index of the \ch{Si3N4} waveguide buried directly underneath.

\subsection{Thermal response characterization through cavity-enhanced photothermal spectroscopy}\label{supp:thermal_response}
For the purpose of programming the pulse sequence to alleviate thermal effects, we study the thermal response of the present transducer due to optical pump heating.
Specifically, thermally induced refractive index change causes a shift in resonance frequency of the \ch{Si3N4} micro-ring, which is detected using cavity-enhanced pump-probe spectroscopy \cite{gaoProbingMaterialAbsorption2022}.
The intensity-modulated pump addresses a TE resonance, while the probe measures the resulting side-of-fringe modulation of another TE mode.
To ascertain the physical mechanism of each cross phase modulation (XPM) process, we sweep the pump modulation frequency to leverage the separation of time scales.
Seen in Supplementary~Fig.~\ref{suppfig:thermal_response}, the response exhibits three plateaus.
We associate the process at modulation frequencies $\gtrsim 1$~MHz with Kerr-induced XPM.
The two slower processes at $\sim 100$~kHz and $\sim 1$~kHz are characteristic of photothermal XPM in quasi-free-standing microresonators such as spheres 
\cite{braginskyOpticalWhisperinggalleryMicroresonators1994} and toroids \cite{schliesserRadiationPressureCooling2006}.
Here, the former ``local'' time scale can be attributed to thermalization of the mode volume with the acoustic resonator, while the thermalization of the suspended structure with the rest of the chip constitutes the latter ``global'' time scale. 
Therefore, it is expected that for future cryogenic operations, employing a pulse-on time ($\tau_\mathrm{on}$) shorter than the local time scale to gate the optical pump will significantly reduce thermal occupancy in the cladding acoustic mode. 
The pulse repetition rate ($f_\mathrm{rep}$) and hence pulse-off time can then be chosen based on the available cooling power.

\section{Characterization of acoustic resonances}\label{supp:acoustics}
The microwave reflection measurements (also known as $S_{11}$) shown here were 
measured at room temperature using custom probes (GGB industries nickel-alloy 40A-GS-135-PC-N Picoprobe) and a vector network analyzer (Rhode\&Schwarz ZNB-20). 
The spectrum shown in Fig.~\ref{fig:fig_2}d is corrected using a calibration substrate (GGB industries CS-8) to remove the phase delay and attenuation introduced by coaxial cables and probes. 
However, this calibration procedure was only used to get the S-parameter, and was not applied during the total conversion efficiency measurements shown in Fig.~\ref{fig:fig_3} and Fig.~\ref{fig:fig_4}.
In a previous version of the transducers and in Ref.~\cite{tianMagneticfreeSiliconNitride2021} where the Band-Aid process (Appendix~\ref{supp:process_flow}) is not employed, the piezoelectric actuator extends beyond the suspended cladding and covers the electrical feedline, causing two detrimental effects for transduction.
First, the resulting stray capacitance leads to excess microwave insertion loss. 
Given the small size of the suspended area ($\sim\SI{1000}{\micro\meter\squared}$) compared to that of the feedline ($\gtrsim\SI{10000}{\micro\meter\squared}$), this effect is readily seen in Supplementary~Fig.~\ref{suppfig:pullback}, where the spectrum without the Band-Aid process exhibits reduced background reflection.
Second, the part of the actuator on top of the unreleased cladding leads to resonant excitation of HBAR modes extending over the entire \ch{Si} substrate. 
These modes have a smaller free spectral range (FSR) of around \SI{19}{\mega\hertz} and a lower microwave extraction efficiency. 
The theory presented in Appendix~\ref{supp:theory} does not correspond to the case of such a multimode system as it assumes being in the sideband-resolved regime, implying the high density of substrate modes is inappropriate for low-noise frequency conversion.
The feedline Band-Aid, discussed in Appendix~\ref{supp:process_flow} and shown in Supplementary~Fig.~\ref{suppfig:pullback}, enables the removal of the piezoelectric layer above the substrate, leaving only HBAR modes well confined in the oxide cladding.

We fit the measured complex reflection to obtain the mechanical quality factor $Q_m = 284$ and microwave extraction efficiency $\kappa_\mathrm{ex, m}/\kappa_m = 0.11$ for the transduction mode \cite{leongPreciseMeasurementsFactor2002}.
This fit model comprises multiple Fano resonances, given by
\begin{equation}
	\label{eq:fano_fit}
	S_{EE}[\omega] = A e^{i \alpha - i\omega \tau} \left( 1 - \sum_r e^{i\phi_r} \frac{2 Q_l^r/Q_c^r }{1 + 2 i Q_l^r (\omega - \omega_r)/\omega_r} \right)
\end{equation}
and the result is shown in Supplementary Fig.~\ref{suppfig:mechanical_modes}a.
In addition, we simulate the microwave reflection of the designed stack geometry using finite-element method (FEM).
Fitting the simulated spectrum leads to $Q_m = 205$ and $\kappa_\mathrm{ex, m}/\kappa_m = 0.61$. 
The discrepancy is attributed to difference in clamping losses and piezoelectric coefficients between the fabricated and simulated devices that lift the degeneracy of electromechanical modes.
We extract the effective mass of the transduction HBAR mode from FEM simulations as
\begin{equation}
	m_\mathrm{eff} = \frac{1}{\max(|\mathbf{x}(\mathbf{r})|^2)} \iiint \rho(\mathbf{r}) \mathbf{x}^*(\mathbf{r}) \mathbf{x}(\mathbf{r}) d\mathcal{V}
	\approx \SI{6}{\nano\gram},
\end{equation}
where $\rho(\mathbf{r})$ is the local mass density and $\mathbf{x}(\mathbf{r})$ is the mechanical displacement, corresponding to $x_\mathrm{ZPF} = \sqrt{\hbar/(2 m_\mathrm{eff}\omega_m)} \approx \SI{2e-8}{\nano\meter}$.

\begin{table*}[hbt!]
	\begin{center}
		\begin{tabular}{|c|c|c|c|c|}
			\hline
			Resonance & $\omega/(2\pi)$ [GHz]& $\kappa/(2\pi)$ [MHz]& $\kappa_\mathrm{ex}/\kappa$ [\%]& $\phi$ [$^\circ$] \\
			\hline
			1 & $\SI{3.479}{}$ & $\SI{13}{}$ & $11$ & $5$ \\
			\hline
			2 & $\SI{3.495}{}$ & $\SI{27}{}$ & $6$ & $5$ \\
			\hline 
			3 & $\SI{3.463}{}$ & $\SI{28}{}$ & $6$ & $11$ \\
			\hline 
			4 & $\SI{3.440}{}$ & $\SI{40}{}$ & $4$ & $-11$ \\
			\hline 
			5 & $\SI{3.397}{}$ & $\SI{38}{}$ & $2$ & $-11$ \\
			\hline 
		\end{tabular}
		\caption{\label{FanoFitTable}Parameters for the Fano resonance fit of the microwave reflection corresponding to eq.(\ref{eq:fano_fit}).}
	\end{center}
\end{table*}

\section{Characterization of bidirectional transduction}\label{supp:parameter_characterization}

\subsection{Estimation of cooperavity and single-photon coupling rate}\label{supp:cooperativity}
Given the measured off-chip efficiency, the single-photon coupling rate can be estimated.
The off-chip efficiency is given by 
\begin{equation}
	\eta^{\mathrm{tot}} = \eta^{\mathrm{probes}} \eta^{\text{fiber-chip}} \eta^{\mathrm{oc}},
\end{equation}
where $\eta_{\mathrm{probes}}$ and $\eta_{\text{fiber-chip}}$ denote the microwave and optical fiber to chip insertion efficiencies, respectively.
The on-chip efficiency can be written as
\begin{equation}
	\eta^{\mathrm{oc}} = \eta^\mathrm{ext}\eta^{\mathrm{int}} = \eta_o \eta_m \eta^{\mathrm{int}}.
\end{equation}
The internal efficiency
\begin{equation}
	\eta^\mathrm{int} = \frac{4C}{(1 \pm C)^2} \approx 4C
\end{equation}
in the low-cooperativity regime, where the plus and minus signs correspond to the anti-Stokes and Stokes configurations, respectively. 
The optical and microwave extraction efficiencies
\begin{align}
	\eta_o &= \frac{\kappa_{\mathrm{ex},o}}{\kappa_o},\\
	\eta_m &= \frac{\kappa_{\mathrm{ex},m}}{\kappa_m}.
\end{align}
The single-photon cooperativity $C_0$ is enhanced by the intracavity photons
\begin{align}
	C &= C_0\bar{n},\\
	n_c 
	&= \kappa_{\mathrm{ex}} |\chi_o|^2 \frac{\eta^\mathrm{fiber-chip}P_{\mathrm{in}}}{\hbar \omega_L}\nonumber\\
	&\approx \frac{4 \kappa_{\mathrm{ex}}}{\kappa_o^2} \frac{\eta^\mathrm{fiber-chip}P_{\mathrm{in}}}{\hbar \omega_L}
	= \eta_o \frac{4}{\kappa_o} \frac{\eta^\mathrm{fiber-chip}P_{\mathrm{in}}}{\hbar \omega_L},
\end{align}
where the pump detuning is assumed to be small. 
The off-chip efficiency with respect to the microwave probe and optical fiber is thus
\begin{align}
	\eta^{\mathrm{tot}} 
	&\approx \eta^{\mathrm{probes}} \eta^{\text{fiber-chip}} \times \eta_m\eta_o \times 4 C_0 \times \eta_o \frac{4}{\kappa_o} \frac{1}{\hbar \omega_L}
	\eta^{\text{fiber-chip}} P_{\mathrm{in}} \nonumber \\
	&= 
	16 \eta^{\mathrm{probes}} \eta^{\text{fiber-fiber}} \times \eta_m \eta_o^2 \times \frac{C_0 P_{\mathrm{in}}}{\hbar \omega_L\kappa_o} \label{eq:off_chip_efficiency} ,
\end{align}
where the fiber-fiber coupling efficiency $\eta^{\text{fiber-fiber}} = \eta^{\text{fiber-chip}}\times\eta^{\text{fiber-chip}}$.
Calibration of the RF probes informed about the probes collections efficiency $\eta_{\mathrm{probes}} \approx \SI{-3}{\decibel}$.
A fraction of light is tapped just before and after the coupling lensed fibers for power monitoring.
We obtain a typical $\eta^{\text{fiber-fiber}} = \SI{-8}{\decibel}$ and hence $\eta^{\text{fiber-chip}} = \SI{-4}{\decibel}$.
Fitting the microwave (optical) reflection spectrum yields $\eta_m = 11\%$ ($\eta_o = 35\%$).
Given that the measured efficiency $\eta^\mathrm{tot} = \SI{-60}{\decibel}$ at $P_{\mathrm{in}} = \SI{10}{\dbm}$, it follows that the single-photon cooperativity is
\begin{equation}
	C_0 = \frac{\eta^{\mathrm{tot}}\times \hbar \omega_L\kappa_o }{16 \eta^{\mathrm{probes}} \eta^{\text{fiber-fiber}} \times \eta_m \eta_o^2 \times P_{\mathrm{in}}}
	\approx \SI{8e-13}{}.
\end{equation}
The single-photon coupling rate between the acoustic and bare optical modes is then
\begin{equation}
	g_0 = 2 \times \sqrt{\frac{\kappa_o \kappa_m C_0}{4}}
	\approx 2\pi\times\SI{42}{\hertz}.
\end{equation}
Here $\kappa_o = 2\pi\times\SI{170}{\mega\hertz}$, $\kappa_m = 2\pi\times\SI{13}{\mega\hertz}$ and the additional factor of two comes from the hybridization of the optical modes (Eq.~\ref{eq:supermode_coupling}).

\subsection{Estimation of on-chip and internal transduction efficiencies}\label{supp:efficiencies}
We have achieved $\eta^\mathrm{tot} = \SI{-48}{\decibel}$ at an input pump power of \SI{21}{\dbm}.
With the losses quoted in Appendix~\ref{supp:cooperativity}, the on-chip and internal efficiencies can be estimated.
For conversion in either directions, the signal goes through the respective input and output ports exactly once, acquiring an attenuation of $\eta^\mathrm{probes}\eta^\mathrm{fiber-chip} = \SI{-7}{\decibel}$.
We therefore have $\eta^\mathrm{oc} = \SI{7.9e-5}{}$.
Knowing the extraction efficiencies $\eta_o = 35\%$ and $\eta_m = 11\%$, we further obtain $\eta^\mathrm{int} = \SI{2e-3}{}$.

\subsection{Non-Lorentzian transduction bandwidth}\label{supp:bandwidth}
\begin{figure*}[htb!]
	\centering
	\includegraphics[width=\columnwidth]{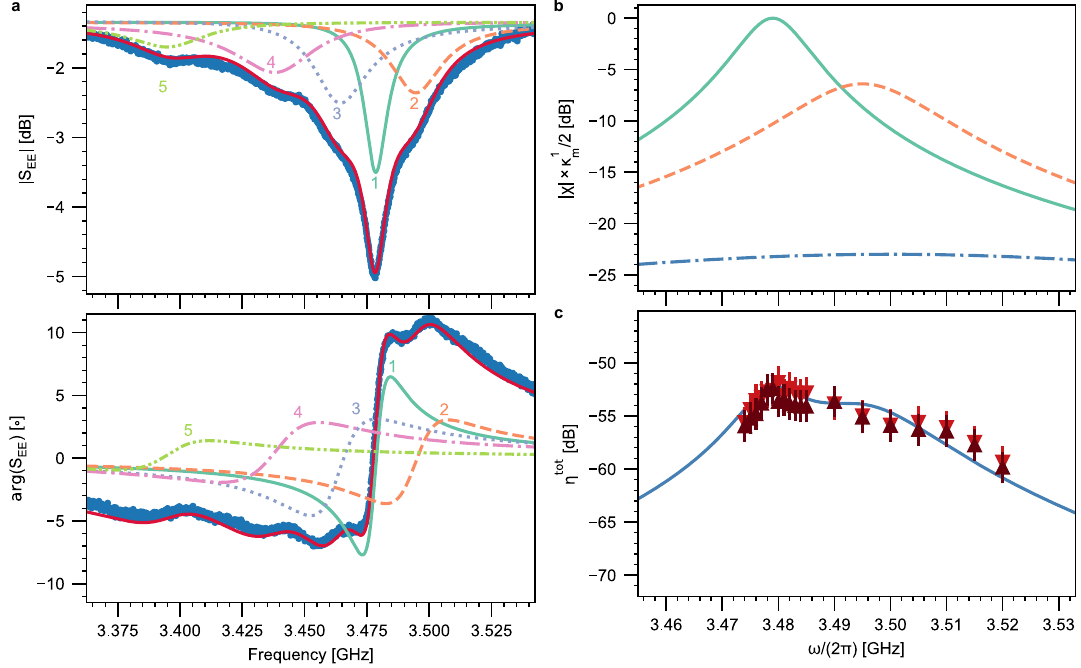}
	\caption{
		\textbf{Spurious acoustic modes near the main resonance.}
		\textbf{a}~Fit of the measured microwave reflection using multiple Fano resonances. 
		The top panel shows the amplitude of the reflection, while the bottom panel depicts its phase after removing the background electrical delay from the cables. 
		The experimental data is shown in blue, the total fit red, and the response of single Fano resonances in the other numbered curves.
		The parameters used for the fit are indicated in Table~\ref{FanoFitTable}.
		\textbf{b}~Susceptibilities of the two acoustic resonances relevant for transduction (solid line for the main resonance, dashed line for the secondary mode), as well as the higher frequency optical mode (dashed-dotted line). The absolute value is normalized to the peak of the main acoustic resonance at \SI{3.479}{\giga\hertz}.
		\textbf{c}~Simulated total transduction efficiency in the anti-Stokes configuration using the two acoustic resonances from \textbf{b}. 
		The triangles correspond to the experimental data from Fig.~\ref{fig:fig_3}b.
		The single photon optomechanical coupling rates are the only free parameters of this fit, which gives $g_0^1 \approx 2\pi \times \SI{20}{\hertz}$ and $g_0^2 \approx 2\pi \times \SI{35}{\hertz}$.
	}
	\label{suppfig:mechanical_modes}
\end{figure*}

The transduction bandwidth shown in Figs.~\ref{fig:fig_3}b and d is not Lorentzian, as should be expected from an ideal transducer in the low cooperativity limit.
We attribute this deviation to the presence of a second mechanical mode contributing to the transduction process.
Fig.~\ref{suppfig:mechanical_modes}b shows the normalized susceptibilities of the principal mechanical mode, the auxiliary mechanical mode as well as the susceptibility of the higher frequency optical mode, considering that the device is operated in the anti-Stokes configuration. 
Fig.~\ref{suppfig:mechanical_modes}c illustrates the transduction spectrum that would be obtained from the theory exposed in Appendix~\ref{supp:theory} considering an additional mode and compares it with the experimental data shown in Fig.~\ref{fig:fig_3}b.
The only free parameters for the fit are the single photon optomechanical coupling rates ($g_0^1 \approx 2\pi \times \SI{20}{\hertz}$ and $g_0^2 \approx 2\pi \times \SI{35}{\hertz}$), the other values being the same as in Appendix~\ref{supp:cooperativity}. 
The difference with the single photon optomechanical coupling rate estimated in Appendix~\ref{supp:cooperativity} comes from the off-resonant contribution of the second mechanical mode.

\section{Optimization of optical extraction efficiency}\label{supp:optimal_coupling}
An optimal optical external coupling can be chosen to maximize the total conversion efficiency.
Equation~\ref{eq:off_chip_efficiency} can be rewritten in terms of quality factors by using the relation $\kappa = \omega/Q$, which yields
\begin{align}
	\eta^{\mathrm{tot}} 
	&\approx 
	16 \eta^{\mathrm{probes}} \eta^{\text{fiber-fiber}} \times \eta_m \eta_o^2  \times g_0^2\frac{Q_m}{\omega_m}\frac{Q_o}{\omega_o} \times \frac{Q_o}{\omega_o}\frac{P_{\mathrm{in}}}{\hbar \omega_L}.
\end{align}
For the optics, we have $Q_o = Q_\mathrm{int} + Q_\mathrm{ext}$, where we separate the intrinsic quality factor $Q_\mathrm{int}$, oftentimes fabrication-limited and hence not easily adjustable, from the external coupling quality factor $Q_\mathrm{ex}$ that can be readily engineered through coupler design.
The optical extraction efficiency can thus be written in the form
\begin{equation}
	\eta_o = \frac{\kappa_{\mathrm{ex}}}{\kappa_{\mathrm{ex}} + \kappa_0}
	= \frac{Q_\mathrm{int}}{Q_\mathrm{ex}} \left(1 + \frac{Q_\mathrm{int}}{Q_\mathrm{ex}}\right)^{-1}
	= \frac{R}{1 + R},
\end{equation}
where $R = Q_\mathrm{int}/Q_\mathrm{ex}$ and the total optical quality factor
\begin{equation}
	Q_o = \frac{1}{Q_\mathrm{int}^{-1} + Q_\mathrm{ex}^{-1}}
	= Q_\mathrm{int} \frac{1}{1 + R}.
\end{equation}
The total efficiency in the low-cooperativity regime is then given by
\begin{equation}
	\eta^{\mathrm{tot}} \approx F \frac{R^2}{(1 + R)^4},
\end{equation}
where 
\begin{equation}
	F = 16 \eta^{\mathrm{probes}} \eta^{\text{fiber-fiber}}  \times \eta_m g_0^2 \frac{Q_m Q_\mathrm{int}^2}{\omega_m\omega_o^2} \times \frac{P_{\mathrm{in}}}{\hbar \omega_L}.
\end{equation}
For a given $F$, the optimal efficiency is achieved at the critical coupling condition $Q_\mathrm{ex} = Q_\mathrm{int}$, or $R = 1$.

\section{Experimental setup and data acquisition}\label{supp:experimental_procedure}
We employ two Toptica CTL 1550 external cavity diode lasers in our experimental setup, depicted in Supplementary Fig.~\ref{suppfig:setup}. 
The first laser (``science laser''), amplified by an erbium-doped fiber amplifier, is used to pump the device and for generating the optical signal for down-conversion, whereas the second laser serves as the local oscillator for optical heterodyne detection. 
The detuning between the two lasers is monitored from their beat note on a photodetector fed to an electronic spectrum analyzer. 
Optical transmission spectra are acquired using an oscilloscope while the science laser is scanned slowly. 
Their frequency axis is calibrated using a Mach-Zehnder interferometer with a long delay line in one arm.
Optical powermeters are inserted before and after the chip, while a variable optical attenuator is used to control the power at the chip input.
Fiber polarization controllers are used to ensure the science laser address the TE modes of the microresonators.

For resonant transduction, the laser is tuned redward into the optical resonance from the blue side \cite{carmon_dynamical_2004}. 
The laser-cavity detuning is chosen to maximize the acousto-optic response, monitored on a vector network analyzer.
Regarding the acousto-optic response, one sideband is absorbed by the optical cavity, while the beating between the unsuppressed sideband and the carrier on a photodiode provides the signal.
For driving and probing the HBAR, the piezoelectric actuator on the chip are connected to high-frequency RF probes (Picoprobe model 40A from GGB industries).
We observed that the microwave reflection spectrum remains stable over the time scale of several months, provided the probe contact condition are maintained. 
We did not observe any change in resonance or broadening within the range of microwave probe power up to 13 dBm and optical pump power up to 25 dBm.
For the down-conversion demonstration, an external electro-optic modulator creates sidebands on the science laser. Their amplitude is adjusted using the optical heterodyne detection setup.
A DC voltage source additionally enables optical mode hybridization using stress-optic effects via the second actuator. 

Pulsed pumping is implemented by driving an acoustic-optic modulator that is switched by a transistor-transistor logic signal generated by an arbitrary waveform generator. 
The pulse sequence is chosen such that the optical resonance does not drift significantly within each pulse, as indicated by a constant steady-state ``pulse-on'' transmission. 
Pulsed up-conversion is recorded utilizing the same heterodyne setup as continuous-wave up-conversion. 
During pulsed down-conversion experiments, the microwave output from the RF probes are routed to a lock-in demodulation setup for detection of weak pulses.

\putbib[references]
\end{bibunit}

\begin{figure*}[htb!]
	\centering
	\includegraphics[width=\columnwidth]{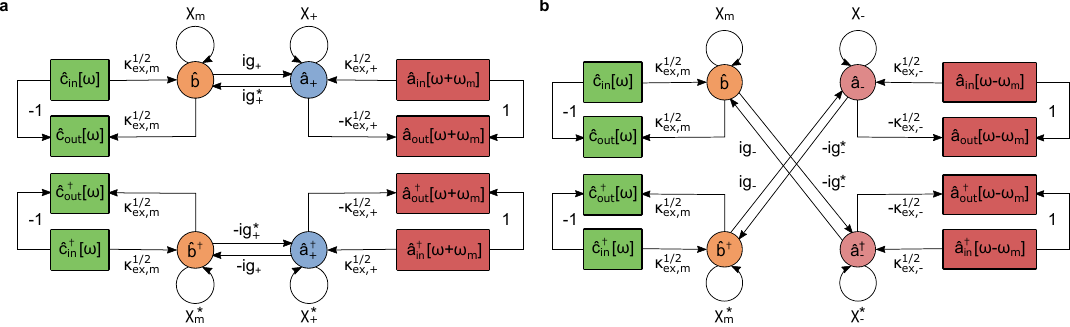}
	\caption{
		\textbf{Signal flow graphs of the frequency conversion processes.}
		\textbf{a}~Effective beam-splitter process.
		\textbf{b}~Effective two-mode-squeezing process.
		The bath operators are indicated by rectangles, the internal modes of the transducers by circles and the coupling rates by arrows.
	}
	\label{suppfig:sfg}
\end{figure*}

\begin{figure*}[htb!]
	\centering
	\includegraphics[width=\columnwidth]{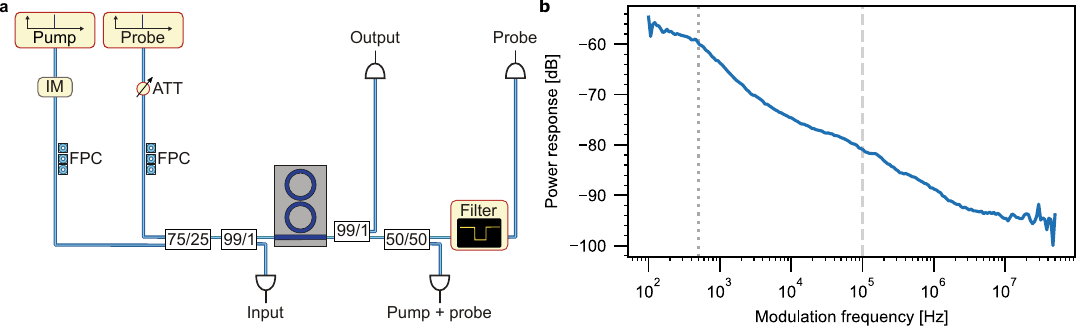}
	\caption{
		\textbf{Thermal response measurement through pump-probe spectroscopy.}
		\textbf{a}~Schematic of the setup used for optical pump-probe spectroscopy. 
		A pump laser, addressing a micro-ring resonance and modulated by an intensity modulator (IM), induces photothermal and Kerr cross-phase modulations.
		As a result, the frequency of a nearby resonance is also modulated, which is detected by a probe laser. 
		The power of the probe laser is adjusted by a variable attenuator (ATT).
		The pump and probe polarizations are optimized with their respective fiber polarization controllers (FPC). 
		The white boxes represent fiber beam splitters.
		\textbf{b}~Measured cross-phase modulation response of a micro-ring as detected at the ``Probe'' photodetector with the pump filtered. 
		The dotted and dashed lines indicate the $3$-dB response bandwidths of the ``global'' and ``local'' photothermal processes, respectively.
	}
	\label{suppfig:thermal_response}
\end{figure*}

\begin{figure*}[htb!]
	\centering
	\includegraphics[width=\columnwidth]{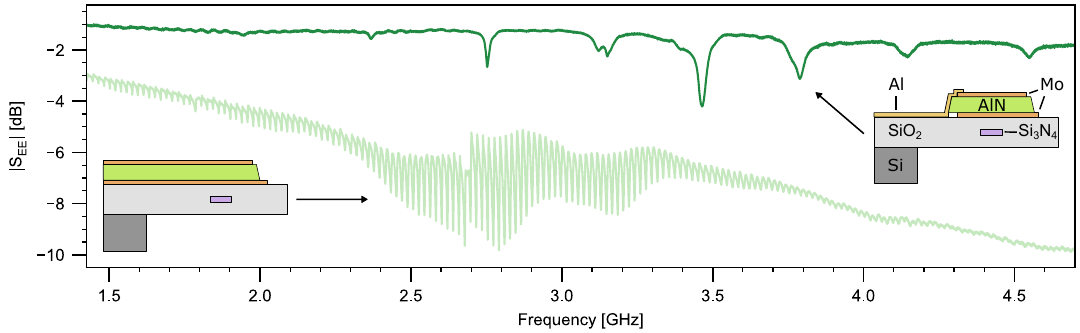}
	\caption{
		\textbf{Elimination of substrate HBAR modes.}
		Microwave reflection spectra obtained with and without applying the Band-Aid process. 
		The respective stack compositions are indicated as insets.
	}
	\label{suppfig:pullback}
\end{figure*}

\begin{figure*}[htb!]
	\centering
	\includegraphics[width=\columnwidth]{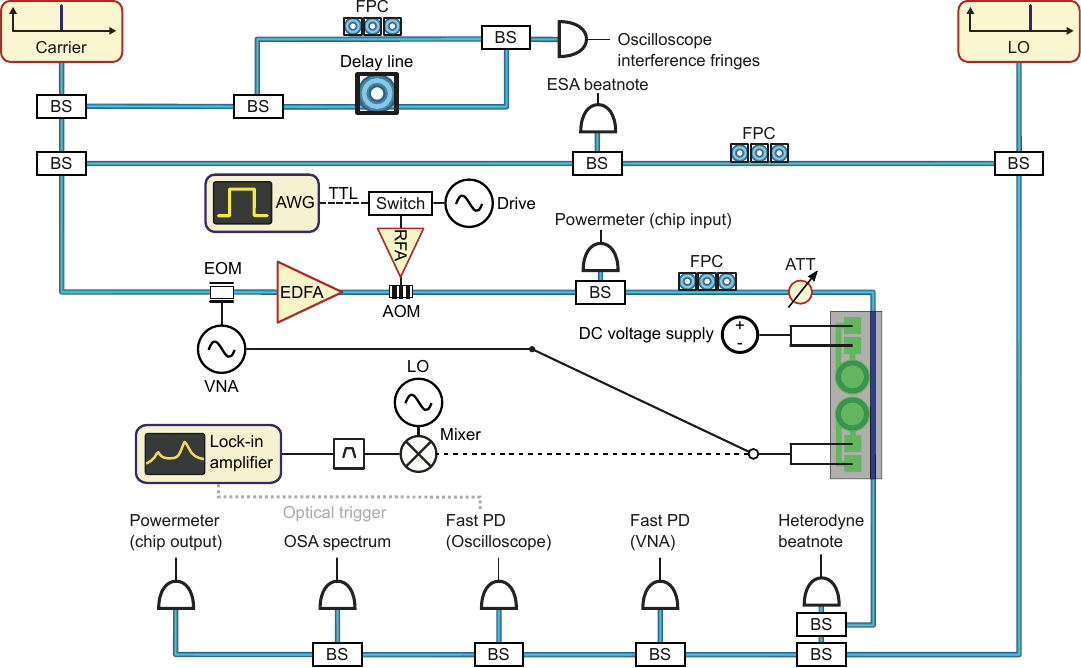}
	\caption{
		\textbf{Experimental setup.}
		The setup to measure the conversion efficiency is divided in three parts. First, part of the laser light is picked to calibrate its frequency using a Mach-Zehnder interferometer where a long delay line is introduced in one of the arm. The second laser, which provide the local oscillator, is beaten with the science laser for frequency calibration of the heterodyne signal. Finally, the science laser goes through the chip. Powermeters are placed before and after the chip in order to measure the insertion loss. The vector network analyzer (VNA) is used during the CW measurements, either for generating optical sidebands or for exciting the mechanics. During the pulsed measurements, the VNA is only used during up-conversion, being replaced by the lock-in detection for the down-conversion. Abbreviations used: AOM: acousto-optic modulator; ATT: variable optical attenuator; AWG: arbitrary waveform generator; BS: beam-splitter; EDFA: erbium-doped fiber amplifier; EOM: electro-optic modulator; FPC: fiber polarization controller; LO: local oscillator; OSA: optical spectrum analyzer; PD: photodetector; RFA: radio frequency amplifier; TTL: transistor-transistor logic.
	}
	\label{suppfig:setup}
\end{figure*}

\begin{figure*}[htb!]
	\centering
	\includegraphics[width=\columnwidth]{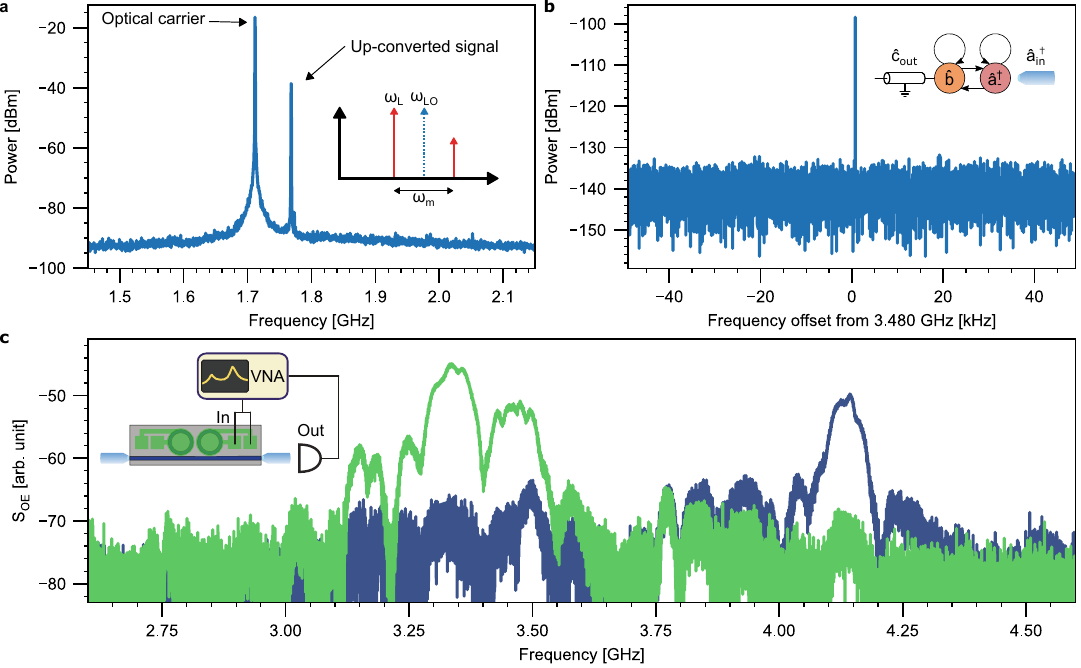}
	\caption{
		\textbf{Experimental dataset examples.}
		\textbf{a}~Optical heterodyne spectrum for a local oscillator set between the carrier and the up-converted sideband, as indicated in the inset.
		\textbf{b}~Microwave probes output measured with an electronic spectrum analyzer. The input optical spectrum contains both the carrier and sidebands generated by an external electro-optic modulator. In this example, the modulation frequency is set to \SI{3.480}{\giga\hertz}.
		\textbf{c}~Acousto-optic response for different laser-cavity detunings on the same transducer.
	}
	\label{suppfig:datasets}
\end{figure*}

\end{document}